\documentclass[twocolumn]{aastex701} %linenumbers,
\newcommand{\ha}{H$\alpha$}
\newcommand{\hb}{H$\beta$}
\newcommand{\paa}{Pa$\alpha$}
\newcommand{\pab}{Pa$\beta$}
\newcommand{\bra}{Br$\alpha$}
\newcommand{\brb}{Br$\beta$}
\newcommand{\oiii}{\mbox{[\ion{O}{3}]}}
\newcommand{\oii}{\mbox{[\ion{O}{2}]}}
\newcommand{\nii}{\mbox{[\ion{N}{2}]}}
\newcommand{\siii}{\mbox{[\ion{S}{3}]}}
\newcommand{\msun}{\ensuremath{M_\odot}}

\newcommand{\smpy}{\ensuremath{M_\odot\,\mathrm{yr}^{-1}}}
\usepackage{multirow}
\usepackage{makecell}
\usepackage{graphicx}
\begin{document}

%\title{COSMOS-3D: A Comprehensive JWST/NIRCam WFSS Census of $\sim12,000$ redshift 0-9 Emission-Line Galaxies}
%\title{A Comprehensive JWST/NIRCam WFSS Census of $\sim12,000$ Emission-Line Galaxies at redshift 0--9\\
%in the COSMOS field}
\title{A Comprehensive JWST Slitless Spectroscopic Census of $\sim$12,000 Emission-Line Galaxies \\
at Redshift $0\sim9$ in COSMOS}

\author[0000-0003-0964-7188]{Shuqi Fu}
\affiliation{Department of Astronomy, School of Physics, Peking University, Beijing 100871, People's Republic of China}
\affiliation{Kavli Institute for Astronomy and Astrophysics, Peking University, Beijing 100871, People's Republic of China}
\affiliation{Jodrell Bank Centre for Astrophysics, University of Manchester, Oxford Road, Manchester UK}
\email{fushuqi@stu.pku.edu.cn}

\author[0000-0002-2420-5022]{Zijian Zhang}
\affiliation{Department of Astronomy, School of Physics, Peking University, Beijing 100871, People's Republic of China}
\affiliation{Kavli Institute for Astronomy and Astrophysics, Peking University, Beijing 100871, People's Republic of China}
\email{zjz.kiaa@stu.pku.edu.cn}

\author[0009-0003-6747-2221]{Danyang Jiang}
\affiliation{Department of Astronomy, School of Physics, Peking University, Beijing 100871, People's Republic of China}
\affiliation{Kavli Institute for Astronomy and Astrophysics, Peking University, Beijing 100871, People's Republic of China}
\email{jiangdy@stu.pku.edu.cn}

\author[0000-0002-4765-1500]{Jie Chen}
\affiliation{Department of Astronomy, School of Physics, Peking University, Beijing 100871, People's Republic of China}
\affiliation{Kavli Institute for Astronomy and Astrophysics, Peking University, Beijing 100871, People's Republic of China}
\email{jiechen@stu.pku.edu.cn}

\author[0000-0003-4176-6486]{Linhua Jiang}
\affiliation{Department of Astronomy, School of Physics, Peking University, Beijing 100871, People's Republic of China}
\affiliation{Kavli Institute for Astronomy and Astrophysics, Peking University, Beijing 100871, People's Republic of China}
\email{jiangKIAA@pku.edu.cn}

\author[0009-0003-2490-4434]{Chenxi Bao}
\affiliation{Department of Astronomy, School of Physics, Peking University, Beijing 100871, People's Republic of China}
\email{2300011608@stu.pku.edu.cn}

\author[0009-0004-6535-9038]{Yihe Wang}
\affiliation{Department of Astronomy, School of Physics, Peking University, Beijing 100871, People's Republic of China}
\email{wangyidaihe@stu.pku.edu.cn}

\author[0000-0002-0322-6131]{Ronaldo Laishram}
\affiliation{National Astronomical Observatory of Japan, 2-21-1 Osawa, Mitaka, Tokyo 181-8588, Japan}
\email{ronaldo.laishram@nao.ac.jp}

\begin{abstract}
Emission-line galaxies provide key diagnostics of star formation, chemical enrichment, and AGN activity across cosmic time. We present a systematic census of emission-line sources using JWST/NIRCam wide-field slitless spectroscopy (WFSS) from the COSMOS-3D program, which covers a large contiguous area of 0.33 deg$^2$ and provides F444W grism spectroscopy over $3.9$--$5.1\ \mu$m. By combining slitless spectroscopy with deep multi-band imaging data, we construct a homogeneous spectroscopic sample of 12,275 objects at redshift 0--9. The typical $3\sigma$ line-flux limit reaches $3.2\times10^{-18}\ {\rm erg\ s^{-1}\ cm^{-2}}$.
To address source confusion inherent to slitless spectroscopy, we develop a multi-step identification framework that combines information from spectra, images, spectral energy distributions, and spatial distributions. We introduce convolutional neural network and large language model to emulate the human identification process and improve efficiency.
The resulting catalog provides spectroscopic redshifts, emission-line measurements, and derived physical properties for all sources. The F444W WFSS enables the detection of multiple emission lines, including primarily \oiii, \ha, \ion{He}{1} $\lambda10830$, \siii\ $\lambda\lambda9531,9069$, \pab, \paa, \brb, PAH, \bra, allowing a continuous census of galaxies across cosmic time.
We construct composite spectra in several redshift bins and detect a variety of faint spectral features that are difficult to identify in individual spectra, including higher-order hydrogen recombination lines, CO absorption bands, and \ion{Fe}{2} lines. This catalog provides a key reference dataset for studies of galaxy evolution and AGN activity, and demonstrates the power of WFSS for constructing large, unbiased spectroscopic samples over wide areas.
The full catalog is publicly available online.
\end{abstract}


\section{Introduction} \label{sec:intro}

Emission-line galaxies play a critical role in studies of galaxy formation and evolution. 
Nebular emission lines provide direct diagnostics of the physical conditions of ionized gas in galaxies, tracing recent star formation, active galactic nucleus (AGN) activity, chemical enrichment, and the physical conditions of ionized gas. These diagnostics make emission-line galaxies powerful probes of galaxy growth across cosmic time. 
Emission lines provide secure spectroscopic confirmation for large samples of high-redshift galaxies \citep{Momcheva_2016, Matthee_2017, Ning2020,Fu_2024, Scholtz_2025, Newman_2025,Pang_2025_GLASS} and have continually pushed the confirmed redshift frontier to ever higher values \citep{Bunker_2023,Castellano_2024,Naidu_2025}.
Large emission-line galaxy samples also enable reliable measurements of the cosmic star formation rate density (SFRD; \citealt{Kennicutt_2012,Madau_2014}). In particular, star formation tracers based on optical and near-IR emission lines are much less affected by dust attenuation than rest-frame UV emission \citep{Ly_2007,Ly_2011,Shioya_2008,Sobral_2013,Bouwens_2021,Kashino_2013,Rinaldi_2023,Bollo_2023,Covelo-Paz_2024,Fu_2025,Korber_2026_O3Hb_LF}. 
In the JWST era, spectroscopic observations of optical and near-IR emission lines at $z>4$ have become available. Over the past three years, JWST observations have revealed systematic evolution between local and high-redshift emission-line galaxies. Such evolution is evident in diagnostic diagrams \citep{Baldwin_1981,Shapley_2015,Shapley_2025,Calabro_2023,HuW_2024}, scaling relations such as the mass--metallicity relation \citep{Curti_2024,Nishigaki_2025}, dust attenuation curves derived from multiple emission lines \citep{Sanders_2025}, increasingly bursty star formation \citep{Clarke_2024,Pirie_2025_JELS}, and the emergence of a large population of broad-line AGNs with properties distinct from those observed in the local Universe \citep{Greene_2024,Matthee_2024}.

Slitless spectroscopy provides an efficient and unbiased way to build such samples. 
The wide-field slitless spectroscopy (WFSS) mode of JWST NIRCam uses grisms in combination with broadband filters to obtain medium-resolution ($R \sim 1600$) spectra over large contiguous areas in parallel imaging fields.  
In particular, WFSS is sensitive to key rest-frame optical to near-IR emission features such as \bra, \brb, PAH, \paa, \pab, \ion{He}{1} $\lambda10830$, \siii, \ha, \hb\ and \oiii\ over $z \sim 0$--9, making it a powerful tool for constructing homogeneous emission-line galaxy samples.
By capturing spectra for all sources in the field of view, it avoids pre-selection and slit losses while producing a continuous census of emission-line systems. 
Different grism orientations can be used to help mitigate spectral overlap, while observations in multiple filters allow the identification of emission lines and extend the accessible redshift range. In practice, the exact combination of grism orientations and filters depends on the observing strategy, and not all configurations are used simultaneously.
JWST NIRCam WFSS programs in Cycles 1--3 have revealed large samples of emission-line galaxies and AGNs out to $z \sim 9$ \citep{Kashino_2023,Meyer_2024,Matthee_2024,2024arXiv240812713S,sun_2025,Sun_2025_SAPPHIRES,Fu_2025}. These surveys have provided the first rest-frame optical spectroscopic view of galaxies and AGNs in the Epoch of Reionization \citep[e.g.,][]{sun_2023,LinX_2024,Oesch_2023,Matthee_2024}, uncovered large-scale structures at $z \sim 5$--9 \citep[e.g.,][]{Champagne_2023, Helton_2024}, and enabled studies of low-mass galaxies that may drive reionization \citep[e.g.,][]{Robertson_2023,Naidu_2024}.

The COSMOS field provides an ideal field for such spectroscopic surveys. It has been one of the most extensively observed extragalactic fields for nearly two decades, with deep and homogeneous imaging from X-ray to radio wavelengths \citep[e.g.,][]{ Scoville_2007,Koekemoer_2007,Steinhardt_2014,Civano_2016, smolcic_2017,Sawicki_2019}. These datasets provide precise photometric redshifts for millions of galaxies \citep[e.g.,][]{Laigle_2016,Weaver_2022} and have supported extensive spectroscopic follow-up with both ground- and space-based facilities \citep[e.g.,][]{Lilly_2007,Brammer_2012,Momcheva_2016,Tasca_2017,Ning2020,vanderWel_2021}.
JWST now provides high-resolution imaging and WFSS coverage across the COSMOS field \citep{Shapley_2025,Malkan_2025}. COSMOS-3D (GO\#5893; PI: K.~Kakiichi) is a large Cycle~3 NIRCam+MIRI imaging and NIRCam WFSS survey whose F444W WFSS observation covers $0.33~\mathrm{deg}^2$ fully within the COSMOS-Web footprint. \citet{Meyer_2025} present a large \oiii\ emitter sample from these data. The combination of continuous slitless spectroscopy and rich ancillary datasets makes COSMOS an exceptional field for constructing an unbiased sample of emission-line galaxies across $0 < z < 9$.

In this work, we present a systematic search for emission-line galaxies in the COSMOS field using NIRCam WFSS observations from the COSMOS-3D program obtained with the F444W filter that covers $3.9$--$5.1\,\mu$m. 
We develop an analysis framework for the NIRCam WFSS data, enabling robust line identification. 
The resulting catalog provides a homogeneous sample of emission-line galaxies with redshifts, measured line fluxes, and derived physical properties, forming a key reference dataset for studies of galaxy evolution and star formation across cosmic time.
This paper is organized as follows. 
Section~\ref{sec:data} describes the observations and data reduction. In Section~\ref{sec:sample}, we detail the sample construction, including emission-line detection, line identification assisted by a large language model (LLM) incorporating spectral, imaging, and photometric information, and line flux measurements. 
Section~\ref{sec:result} presents the main results, including the final line-emitter catalog, composite spectra, and physical properties. 
In Section~\ref{sec:discuss}, we discuss systematic effects and interpretation. 
Finally, we summarize our main conclusions in Section~\ref{sec:summary}.

Throughout this paper, we assume a flat $\Lambda$CDM cosmology with $H_0=70\ \mathrm{km}\ \mathrm{s}^{-1}\ \mathrm{Mpc}^{-1}$ and $\Omega_m = 0.3$. The AB magnitude system \citep{Oke_1983} is used. 
%We also assume a \citet{Chabrier_2003} initial mass function.

\section{Observations and Data Reduction} \label{sec:data}
In this section, we describe the direct imaging and grism datasets used in our analysis and the procedures adopted for their reduction. We first summarize the assembly and processing of the multi-band direct imaging data and the construction of the photometric catalog. We then outline the reduction and extraction of the NIRCam grism spectroscopy used in this work.

\subsection{Direct Imaging} \label{sec:image}

We make use of the complete set of JWST/NIRCam imaging available in the COSMOS field, primarily including: (1) the 0.54~deg$^{2}$ four-band imaging (F115W, F150W, F277W, F444W) from \textit{COSMOS-Web} (GO\#1727; \citealt{Cweb_survey_ref}; PIs: Kartaltepe and Casey); (2) the eight-band imaging (F090W, F115W, F150W, F200W, F277W, F356W, and F410M, F444W) from \textit{PRIMER} (GO\#1837; PIs: Dunlop) in the CANDELS region; and (3) the 0.33~deg$^{2}$ F115W, F200W, and F356W imaging from \textit{COSMOS-3D} (GO\#5893; PIs: Kakiichi, Egami, Fan, Lyu, Wang, and Yang). Other smaller programs and parallel observations are also included, as detailed in Table \ref{tab:img_data}. These additional programs provide coverage for new bands and deep exposures in certain regions. The total exposure time of the imaging data is 810.3 hours.

We reduced the NIRCam imaging data following the steps in our previous works \citep{2025ApJ...985..119Z, Liu_2026,Zhang2026}. We reduced the raw data using the combination of the \textit{JWST Calibration Pipeline} (v.1.17.0) with CRDS version 11.17.0 and CRDS context 1185, the CEERS \citep[][]{2023ApJ...946L..12B} NIRCam imaging reduction\footnote{\url{https://github.com/ceers/ceers-nircam.}} and our own custom codes. Here, we provide a brief description of the steps.

Stage 1 of the JWST Calibration Pipeline performs detector-level corrections, starting from {\tt\string uncal.fits} files and ending with {\tt\string rate.fits} files in units of count/s. Following the CEERS team's scripts, before running \hbox{ramp-fitting} step in Stage 1 of the pipeline, we identified snowballs, increased their footprints, and flagged them as {\tt\string JUMP\_DET} so that snowballs were  removed after the ramp-fitting step. ``Wisp'' features from short-wavelength (SW) images were subtracted using the latest official wisp templates\footnote{\url{https://stsci.app.box.com/s/1bymvf1lkrqbdn9rnkluzqk30e8o2bne.}}.
The 1/f noise, i.e., horizontal and vertical striping patterns in the images, were subtracted with {\tt\string remstriping.py} in the CEERS team's scripts.

We then ran Stage 2 of the JWST Calibration Pipeline with the default parameters, which involves individual image calibrations such as flat-fielding and flux calibration. Astrometry was calibrated using {\tt\string TweakregStep}, with {\tt\string abs\_refcat} set to a source catalog detected from the HST COSMOS F814W images\footnote{\url{https://irsa.ipac.caltech.edu/data/COSMOS/images/acs_mosaic_2.0/}.}.
After running {\tt\string SkyMatchStep} and {\tt\string OutlierDetectionStep}, we subtracted a 2D background for each image following the method described in \citet{2023ApJ...946L..12B}, and then ran {\tt\string ResampleStep} to drizzle and combine images to make mosaic images. 
We set {\tt\string pixfrac} to 0.8 and pixel scale to $0\farcs03$ for all filters. 
{\tt\string CRVAL} and {\tt\string CRPIX} are the same for each filter so that all mosaics have the same WCS grid. 
Finally, a 2D background was subtracted from each mosaic using the method described in \citet{2023ApJ...946L..12B}.

\begin{table*}
\caption{Program information for the imaging data used in this work, including primary and parallel observations.}
\label{tab:img_data}
\centering
\setlength{\tabcolsep}{2pt} 
\begin{tabular}{cccc@{\hspace{-25pt}}cllc}
% \toprule
\hline\hline
PROGRAM & $t_{\mathrm{exp}}^{\mathrm{eff}}$ (h) & $N_{\mathrm{visits}}$ & $N_{\mathrm{filters}}$ & Filters & Obs. start & Obs. end  & Reference\\
% \midrule
\hline
1635 & 45.2 & 15 & 7 & \makecell[tl]{F090W, F115W, F150W, F150W2, \\F322W2  F410M, F444W} & 2022-12-04 & 2023-05-21 \\
\hline
1727& 176.4 & 242 & 4 & \makecell[tl]{F115W, F150W, F277W, F444W} & 2023-01-05 & 2024-05-17 & \citet[][]{Cweb_survey_ref}\\
 (COSMOS-Web)  \\
\hline
\makecell[tc]{\hspace{-35pt}1810 \\ \hspace{-35pt}(Blue Jay)} & 67.2 & 3 & 7 & \makecell[tl]{F090W, F115W, F150W, F200W,  \\F277W F356W, F444W} & 2022-11-25 & 2022-12-27 & \citet{BlueJay_Belli2025} \\
\hline
\makecell[tc]{\hspace{-35pt}1837 \\ \hspace{-35pt}(PRIMER)} & 97.9 & 98 & 8 & \makecell[tl]{F090W, F115W, F150W, F200W, \\F277W  F356W, F410M, F444W} & 2022-12-26 & 2024-05-21 \\

\hline
1840 & 2.1 & 3 & 8 & \makecell[tl]{F115W, F150W, F200W, F250M, \\F277W  F335M, F356W, F444W} & 2022-12-20 & 2023-04-25 \\
\hline
1933 & 12.9 & 6 & 4 & \makecell[tl]{F115W, F150W, F200W, F356W} & 2023-01-06 & 2023-01-07 \\
\hline
\makecell[tc]{\hspace{-31pt}2321 \\ \hspace{-31pt}(JELS)} & 14.0 & 4 & 2 & \makecell[tl]{F212N, F470N} & 2024-11-23 & 2024-11-24 & \citet{Duncan2025_JELS}\\
\hline
2362 & 1.3 & 3 & 2 & \makecell[tl]{F200W, F444W} & 2023-04-25 & 2023-05-06 \\
\hline
2417 & 0.7 & 1 & 3 & \makecell[tl]{F200W, F356W, F444W} & 2024-04-05 & 2024-04-05 \\
\hline
\makecell[tc]{\hspace{-35pt}2514 \\ \hspace{-35pt}(PANORAMIC)}  & 17.0 & 24 & 6 & \makecell[tl]{F115W, F150W, F200W, F277W, \\F356W  F444W} & 2023-12-02 & 2024-01-06 & \citet{Williams2025_PANORAMIC}\\
\hline
2775 & 3.7 & 2 & 5 & \makecell[tl]{F070W, F115W, F356W, F430M, \\F460M} & 2024-05-19 & 2024-05-19 \\
\hline
\makecell[tc]{\hspace{-35pt}3990 \\ \hspace{-35pt}(BEACON)} & 12.6 & 5 & 8 & \makecell[tl]{F090W, F115W, F150W, F200W, \\F277W  F356W, F410M, F444W} & 2024-05-14 & 2024-05-24 & \citet{BEACON2025}\\
\hline
4265 & 7.3 & 3 & 7 & \makecell[tl]{F090W, F115W, F150W, F200W, \\F277W  F356W, F444W} & 2024-01-04 & 2024-04-23 \\
\hline
5324 & 1.3 & 2 & 6 & \makecell[tl]{F115W, F150W, F200W, F277W, \\F356W  F444W} & 2025-05-06 & 2025-05-19 \\
\hline
\makecell[tc]{\hspace{-35pt}5398 \\ \hspace{-35pt}(POPPIES)} & 26.1 & 13 & 8 & \makecell[tl]{F070W, F090W, F115W, F150W, \\F200W  F277W, F356W, F444W} & 2025-04-17 & 2025-05-04 \\
\hline
\makecell[tc]{\hspace{-35pt}5893 \\ \hspace{-35pt}(COSMOS-3D)} & 166.3 & 187 & 3 & \makecell[tl]{F115W, F200W, F356W} & 2024-11-24 & 2025-12-21 & \makecell[tc]{\hspace{-35pt}\citet{Meyer_2025} \\ \hspace{-35pt}\citet{Lin2026_C3D}}\\
\hline
\makecell[tc]{\hspace{-35pt}6368 \\ \hspace{-35pt}(CAPERS)} & 63.0 & 9 & 12 & \makecell[tl]{F115W, F140M, F150W, F182M, \\F200W  F210M, F250M, F277W, \\F335M, F356W  F444W, F480M} & 2025-04-13 & 2025-05-22 \\
\hline
\makecell[tc]{\hspace{-35pt}6434 \\ \hspace{-35pt}(SAPPHIRES)} & 75.0 & 113 & 12 & \makecell[tl]{F070W, F090W, F115W, F150W, \\F182M  F200W, F210M, F277W, \\F335M, F356W  F410M, F444W} & 2025-04-05 & 2025-05-25 & \citet{Sun_2025_SAPPHIRES}\\
\hline
6480 & 7.1 & 7 & 7 & \makecell[tl]{F070W, F090W, F115W, F150W, \\F200W  F360M, F444W} & 2025-04-07 & 2025-05-24 \\
\hline
6585 & 11.3 & 3 & 4 & \makecell[tl]{F115W, F150W, F277W, F444W} & 2024-04-29 & 2024-04-29 \\
\hline
9372 & 2.1 & 2 & 5 & \makecell[tl]{F150W, F200W, F277W, F356W, \\F444W} & 2025-05-27 & 2025-12-13 \\
\hline
% \bottomrule
\end{tabular}
\end{table*}

\begin{figure*}[t]
\centering
\includegraphics[width=\textwidth]{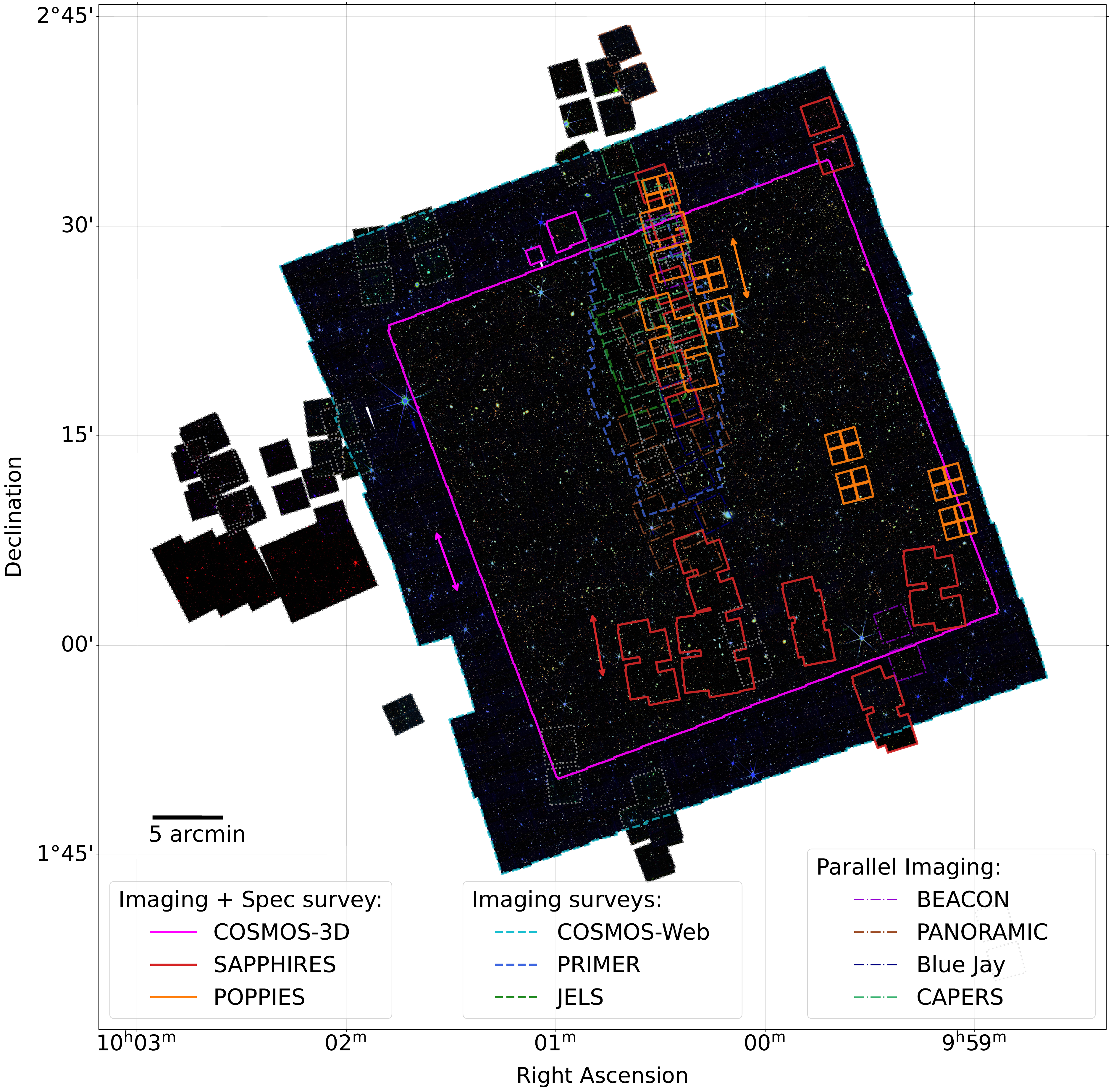}
\caption{JWST/NIRCam false-color image of the mosaics including all public data in the COSMOS field. This image is made using the six bands that cover the whole field. The following false-color coding was adopted: blue - F115W + F150W; green - F200W + F277W; red - F356W + F444W. North is up, east is left. Region covered by major surveys (e.g., total exposure time $>$ 12 h) are labeled with color-coded lines, while dotted gray lines show the region with parallel observations from other smaller programs. Surveys with slitless spectroscopic observations are enclosed by solid lines, while the dispersion direction is marked by arrows.
}
\label{fig:cover}
\end{figure*}

\subsection{Photometric Catalog} \label{sec:phot_cat}

We constructed a new multi-band photometric catalog based on the full set of reduced JWST/NIRCam imaging described above, in order to obtain consistent photometry for our photometric analyses. The catalog incorporates not only the COSMOS-Web imaging in the F115W, F150W, F277W, and F444W bands, but also the additional F200W and F356W imaging from COSMOS-3D and other JWST programs (Figure~\ref{fig:cover}). These additional bands are particularly valuable for improving and validating the photometric redshifts of high-redshift galaxy candidates.

The COSMOS2025 catalog \citep{COSMOS2025_cat} provides photometry, morphology, photometric redshifts, and physical parameters for more than 700,000 objects in the COSMOS field. 
The detection images are constructed from the deep JWST/NIRCam imaging in the F115W, F150W, F277W, and F444W bands obtained by the COSMOS-Web program, with the F444W imaging reaching the current maximum depth available for source detection over most of the COSMOS field. 
This catalog is therefore well suited for use in NIRCam F444W WFSS extraction within the COSMOS-3D footprint. 
However, COSMOS2025 does not incorporate the COSMOS-3D imaging, particularly the additional F200W and F356W bands, nor imaging from other JWST programs.
We therefore employ the COSMOS2025 photometry catalog as the primary catalog for the NIRCam F444W WFSS extraction (Section \ref{subsec:line_detect}). 
We match our newly constructed photometric catalog to COSMOS2025 and retain the COSMOS2025 source IDs for matched objects. In addition, we supplement the extraction catalog with bright ($m_{\rm F444W}<24$) sources detected in our
catalog that do not have COSMOS2025 counterparts, as described below.
The newly measured multi-band photometry is also used for photometric
redshift estimation (Section \ref{subsec:photz}) and SED modeling (Section \ref{sec:sed}).
%while adopting our own newly constructed photometry (see below) for photometric redshift measurements, SED modeling, etc.

%The COSMOS2025 catalog \citep{COSMOS2025_cat} provides photometry, morphology, photometric redshifts, and physical parameters for more than 700,000 objects in the COSMOS field. The detection images are constructed from the deep JWST/NIRCam imaging in the F115W, F150W, F277W, and F444W bands obtained by the COSMOS-Web program, with the F444W imaging reaching the current maximum depth available for source detection over most of the COSMOS field.
%currently set the limiting depth for direct detection in this band. 
%This catalog is therefore well suited for use in NIRCam F444W WFSS extraction within the COSMOS-3D footprint. However, COSMOS2025 does not incorporate the COSMOS-3D imaging, particularly the additional F200W and F356W bands, nor imaging from other JWST programs.
%These additional data are especially valuable for improving and validating the photometric redshifts of high-redshift galaxy candidates.
%We therefore constructed a new catalog based on the full set of reduced NIRCam imaging described above to obtain consistent multi-band photometry for our photometric analyses. 

% spike removal
% PSF construction and convolution,detection image
% source detection and aperture correction 
% HSC data reprocessing
% HSC PSF construction and tphot
% Error estimation
% photometric estimation
%\subsubsection{Spike Masking}
\subsubsection{Spike Masking, PSF Matching, and Detection Image Construction}
\begin{figure*}[t]
\centering
\includegraphics[width=0.9\textwidth]{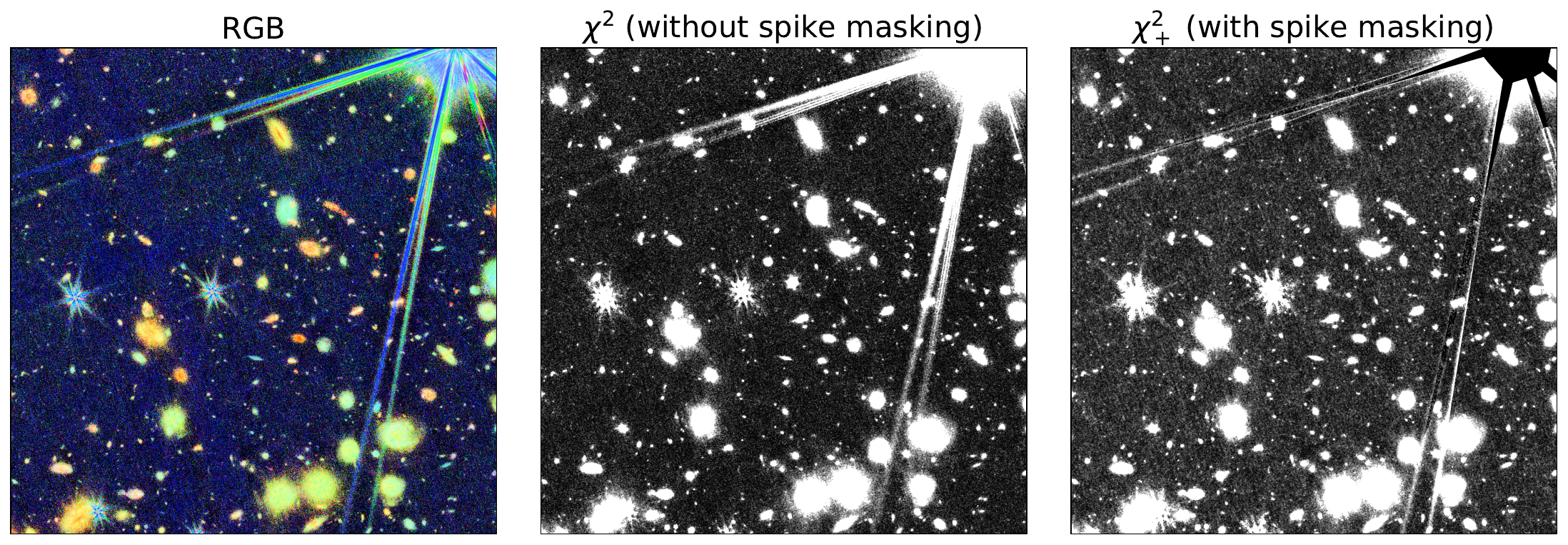}
\caption{Comparison between the standard $\chi^2$ detection image and the spike-masked $\chi^2_+$ image. Left: JWST RGB composite image constructed from F115W+F150W (blue), F200W+F277W (green), and F356W+F444W (red). Middle: original $\chi^2$ image without diffraction-spike masking. Right: $\chi^2_+$ image after applying spike masking. 
\label{fig:spike_demonstration}}
\end{figure*}

%In the JWST imaging data, some bright stars produce diffraction spikes that can contaminate nearby sources. 
Before constructing the detection image, we masked these bright spikes to avoid their imprint in the detection process. We did not rely on automated PSF template masking. Instead, we manually masked the spikes to remove the contamination while preserving genuine sources. Due to the fact that different bands sometimes have different observational position angles, sources masked in one band may not be masked in others, and thus can still appear in the detection image. This is demonstrated in the right panel of Figure~\ref{fig:spike_demonstration}. The masking procedure effectively suppresses spurious high-significance structures associated with diffraction spikes and detector artifacts around bright stars.

%\subsubsection{PSF Construction, Image Convolution, and Detection Image}
For the JWST NIRCam and HST images, we constructed empirical PSF models for each band using \texttt{PSFEx} \citep{2011ASPC..442..435B}, following the method in \citet{2024ApJ...962..139Z} with slight modifications. 
Briefly, we selected high signal-to-noise ratio (SNR; \texttt{SNR\_WIN} $>100$) candidate PSF sources that are non-blended (\texttt{FLAGS} $< 2$), non-irregular (\texttt{ELONGATION} $< 1.5$), point-like (\texttt{CLASS\_STAR} $>0.8$), and without bad pixels (\texttt{IMAFLAGS\_ISO}~$=0$).
% \textcolor{red}{(need check)}
We adopted a third-order polynomial with ten vector images to model the spatial variations of the PSF model across the field of view as a function of pixel coordinates (x and y).
The PSF models were constructed with an oversampling=2. 
The final image size of PSF models is 265$\times$265 pixels for SW filters and 401$\times$401 pixels for LW filters.

\begin{figure}[!t]
\centering
\includegraphics[width=0.5\textwidth]{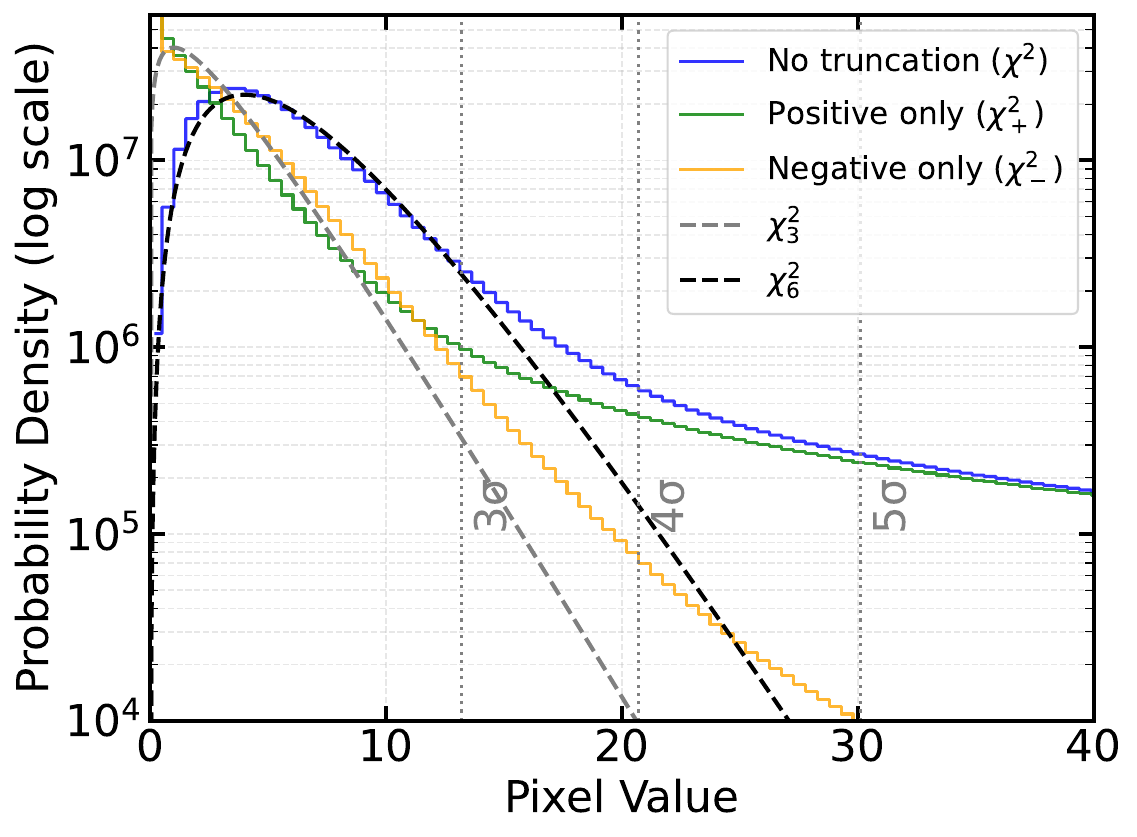}
\caption{Statistical distribution of pixel values of the $\chi^2$ images described in Section \ref{subsubsec:photometry}. The blue curve represents the full distribution, while the green and yellow histograms indicate the positive ($\chi^2_+$) and negative ($\chi^2_-$) truncated distributions, respectively. The $\chi^2_+$ maps are used as detection images. Theoretical $\chi^2$ profiles for degrees of freedom $N=6$ (dark grey dashed) and $N=3$ (light grey dashed) are overlaid for comparison. Vertical dashed lines denote the $3\sigma$, $4\sigma$, and $5\sigma$ significance levels relative to an $N=3$ $\chi^2$ distribution.\label{fig:chi_distribution}}
\end{figure}

We used these PSFs to derive kernels to match all PSFs to the F444W PSF using \texttt{PyPHER} \citep{2016ascl.soft09022B}, and then used them to convolve all images to match the F444W image. 
The detection image is an combination of the PSF-matched JWST images of the six bands that provide nearly full coverage of the fields (F115W, F150W, F200W, F277W, F356W, and F444W) using a $\chi^2$ technique \citep{1999AJ....117...68S}.
The $\chi^2$ image was constructed following the method in \citet{COSMOS2025_cat}, which performs a truncation (setting negative pixels to zero) of the individual SNR maps before combination in order to suppress spurious detections.
Because not all pixels are covered by all six bands due to spike masking and the relatively small coverage of the F200W and F356W bands from COSMOS-3D, the effective number of degrees of freedom (dof) varies across the image. 
To account for this, we converted the $\chi^2$ value at each pixel to a $p$-value using its local dof, and then rescaled it to an equivalent $\chi^2$ value with a reference dof = 6. 
The final detection image is defined as the square root of this rescaled $\chi^2$ map, ensuring a uniform statistical interpretation across the field. The $\chi^2$ distribution for this data set is shown in Figure~\ref{fig:chi_distribution} and an example $\chi^2_+$ image is shown in the middle panel of Figure~\ref{fig:spike_demonstration}. The observed $\chi^2$ distributions are broadly consistent with the expected theoretical $\chi^2$ behavior.

\subsubsection{Photometry of JWST and HST Images}
\label{subsubsec:photometry}
We performed photometry using \texttt{SExtractor} \citep{Bertin_1996} on the PSF-matched images to construct the source catalog. 
We used the dual image mode using the detection image constructed above. 
For the individual band images used for photometric measurements, no spike masking was applied to the science images themselves.
Nevertheless, spike masks were retained as separate mask maps on a band-by-band basis.
For sources whose photometric apertures overlap with spike-masked regions in a given band, the flux measurement in that band is flagged, while photometry in other bands is preserved.
We adopt a two-step `cold + hot' mode to achieve optimal source detection \citep{2012MNRAS.422..449B}, with detailed parameters listed in Table \ref{tab:sex_params} of the Appendix.
% Photometry is measured in a small Kron aperture ($k$=1.3, $R_{\min}$=1.8), and an aperture correction is performed to match the default Kron aperture ($k$=2.5, $R_{\min}$=3.5) in each band \textbf{(more details)}. 

We initially measured photometry using a small Kron aperture ($k=1.3$, $R_{\min}=1.8$). 
We then applied aperture corrections to match the default Kron aperture ($k=2.5$, $R_{\min}=3.5$) while accounting for variations in source morphology.
The correction was parameterized as a function of both source size and compactness, quantified by the Kron radius and the flux ratio between two circular apertures ($F_{0\farcs25}/F_{0\farcs5}$), respectively. 
We divided sources into quantile bins in this two-dimensional parameter space to ensure sufficient statistics in each bin.
Within each bin, the median flux ratio between the large and small Kron apertures is computed in magnitude intervals and modeled as a linear interpolation function of magnitude. 
For bins with insufficient sources, we adopted a robust global fit as a fallback. We performed this procedure independently in each band.

Following the method in \citet{2023ApJS..269...16R}, we measured the photometric errors by placing 100,000 random apertures in the source-free region of the images. These apertures were divided into 100 groups based on their weight values. For each group, a power-law relation between aperture size and flux scatter was fitted and used to derive the sky background noise for each source according to its weight value and aperture size. The final error estimates account for both sky background noise and Poisson noise.

\begin{table}[t!]
\begin{center}
   \caption{Relevant \texttt{SExtractor} parameters for the cold + hot aperture photometry in Section \ref{subsubsec:photometry}.\label{tab:sex_params}}
   \hspace*{-0.5cm}
   \begin{tabular}{lll}
\hline\hline
Parameter                 & hot      & cold      \\[4pt]
\hline
% \texttt{PHOT\_APERTURES}             & max($4\times$FWHM, 5 pixels)  &            \\
\texttt{PHOT\_AUTOPARAMS}             & \multicolumn{2}{c}{2.5, 3.5}      \\
\texttt{DETECT\_MINAREA}  & 5   &      15                           \\
\texttt{DETECT\_THRESH}   & 3.64   &   4.55                   \\
\texttt{ANALYSIS\_THRESH} & 3.64   &   4.55                      \\
\texttt{DEBLEND\_NTHRESH} & \multicolumn{2}{c}{64}         \\
\texttt{DEBLEND\_MINCONT} & 0.0005   &   0.001                 \\
\texttt{FILTER\_NAME}     & gauss\_3.0\_7x7   &    tophat\_9.0\_9x9         \\
\texttt{GAIN}             & \multicolumn{2}{c}{0}               \\
\texttt{BACK\_TYPE}       & \multicolumn{2}{c}{MANUAL}          \\
\texttt{BACK\_VALUE}      & \multicolumn{2}{c}{0}      \\
\texttt{BACK\_SIZE}       & 2000          &     4000        \\
\texttt{BACK\_FILTERSIZE} & \multicolumn{2}{c}{3}            \\
\texttt{BACKPHOTO\_THICK} & \multicolumn{2}{c}{120}         \\
\hline
   \end{tabular}
\end{center}
\begin{minipage}{0.485\textwidth}{\small}

\end{minipage}
\end{table}

\subsubsection{Photometry of Ground-based Images}
We also used Ultradeep optical ($grizy$) imaging data from the Hyper Suprime-Cam Subaru Strategic Program \cite{2022PASJ...74..247A} third public data release (PDR3) and the CFHT $U$-band images from the MUSUBI (MegaCam Ultra-deep Survey: u*-band Imaging) survey \cite{2022ApJS..260...54W}. These deep optical data provide important constrains on the rest-frame UV continuum and Lyman-break features for high-redshift galaxy candidates, thereby improving the accuracy and robustness of photometric redshift estimates.
For the HSC and CFHT images, we derived template-fitting deblended photometry to obtain total fluxes at each band using T-PHOT v2.0 \citep{2016A&A...595A..97M}. T-PHOT
uses position priors from the high-resolution NIRCam F444W images. We generated PSF models for the NIRCam, HSC, and CFHT images using \texttt{PSFEx}, and constructed convolution kernels to match the NIRCam F444W PSF to those of the HSC and CFHT images. These kernels were used to build image templates for each source, from which the photometry in each band is extracted through template fitting with T-PHOT.

%\medskip
Finally, we matched our catalog to the COSMOS2025 catalog using a $0\farcs5$ radius. We found that the bright-star masking in COSMOS2025 removes some potential bright-line emitters, which may lead to incorrect line identification for a subset of detected emission lines. We therefore retained sources with F444W band photometry brighter than 24 mag in our catalog that do not have COSMOS2025 counterparts, and we visually inspected them and removed spurious detections. 
These additional sources were included in the line-emitter identification step to reduce misidentifications, but we recommend excluding them from statistical analyses such as luminosity-function measurements because the corresponding masked area is not accounted for and their photometric redshifts may be unreliable due to contamination. 
The final photometric catalog used for spectral extraction and emission-line identification adopts the same source IDs as the COSMOS2025 catalog. Additional sources without COSMOS2025 counterparts are assigned IDs larger than 1,000,000.

\begin{figure*}[t]
\centering
\includegraphics[width=\textwidth]{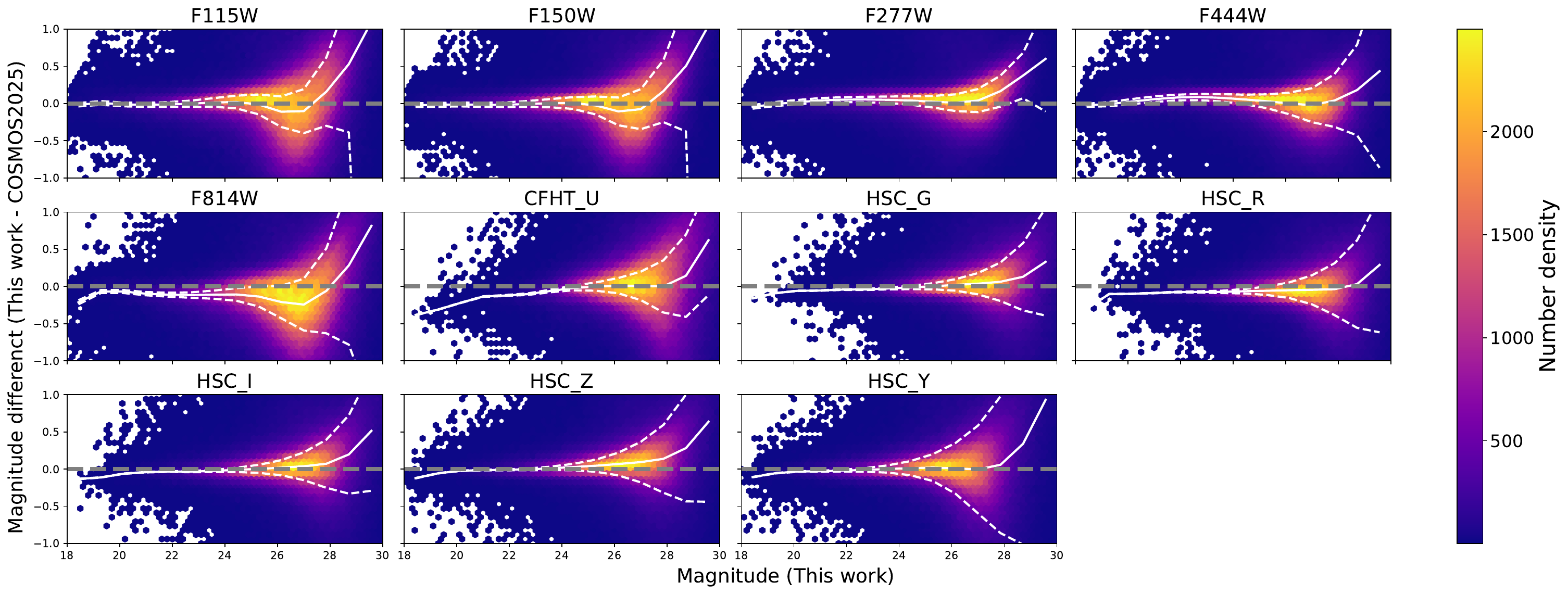}
\caption{Comparison of photometric measurements between the COSMOS2025 catalog and our catalog in the JWST F115W, F150W, F277W, and F444W bands, the HST F814W band, the CFHT $U$ band, and the HSC $g$, $r$, $i$, $z$, and $y$ bands. The panels show the magnitude difference ($m_{\mathrm{This~work}}-m_{\mathrm{COSMOS2025}}$) as a function of the measured magnitude in this work. The color scale indicates the source number density. The solid white curves represent the median magnitude offset in bins of magnitude, while the dashed white curves show the $16^{\mathrm{th}}$ and $84^{\mathrm{th}}$ percentiles. The horizontal dashed gray line marks zero magnitude difference. Overall, the two catalogs show good agreement over a wide magnitude range, with increasing scatter toward fainter magnitudes.  \label{fig:phot_compare_cos25}
}
\end{figure*}

\subsection{Grism Spectroscopy} \label{sec:spec}

COSMOS-3D obtained NIRCam WFSS observations of the COSMOS field using the row-direction grism (Grism~R) and the F444W wide-band filter. 
We reduced the NIRCam WFSS data using a customized pipeline based on the procedures described by \citet[e.g.][]{sun_2023,Sun_2025_SAPPHIRES}. The corresponding codes and calibration files are publicly available\footnote{\url{https://github.com/fengwusun/nircam_grism}\label{fws_script}}.

We processed the grism exposures with the standard JWST Stage-1 calibration pipeline (version \verb|1.13.4|) and the calibration reference file \verb|jwst_1364.pmap|. We then assigned WCS information to the grism images and applied flat-field corrections using calibration flats obtained with the same filter and detector configuration. 
For each exposure, we constructed a super-sky background frame from all flat-fielded images with the same filter+pupil+module configuration and subtracted it. Residual background structures remained after this step, so we performed an additional two-dimensional sky-background subtraction using the Source Extractor algorithm \citep{Bertin_1996}. We estimated the background from a $2.5\sigma$-clipped image using the mode estimator $(2.5 \times \mathrm{median}) - (1.5 \times \mathrm{mean})$, which is less sensitive to source crowding. When 
$(\mathrm{mean} - \mathrm{median})/\mathrm{STD} > 0.3$,
where $\mathrm{STD}$ is the standard deviation, we considered the mode estimator unreliable and instead adopted the median value. 
We further removed residual $1/f$ noise by subtracting the median background along both row and column directions of the grism images. We performed this correction iteratively on continuum-subtracted emission-line maps, while masking high-significance pixels to avoid over-subtraction around bright emission features. Both the original grism images and the continuum-subtracted emission-line images produced by median filtering along the dispersion direction are saved. 

Next, we corrected the astrometry of the LW grism images using the simultaneously observed SW direct images. We reduced the SW images with the JWST stage-2 imaging pipeline and detected compact sources using \texttt{DAOStarFinder}. We then matched these detections to the COSMOS2025 source catalog, which is registered to the \textit{HST} astrometric reference frame, to measure the astrometric offsets and rotation for each exposure. We applied these corrections to the LW grism WCS, since even small astrometric offsets can significantly affect the spectral trace, wavelength solution, and flux calibration of NIRCam WFSS spectra.
We then generated source catalogs for each LW grism exposure by projecting the master photometric catalog onto the astrometry-corrected grism WCS and identifying sources that could produce spectra in each exposure. 

Finally, we extracted the 2D spectra. For each source and exposure, we used the calibrated spectral tracing and dispersion functions %\footref{fws_script} 
to determine the wavelength-dependent trace position on the detector. We extracted both the 2D spectra and calibrated their fluxes using the module- and grism-dependent sensitivity curves. We then resampled individual spectra from different exposures onto a common wavelength grid with $\Delta\lambda = 1$~nm and coadded them to produce the final 2D spectra for each source. The default height of the extracted 2D spectra in the spatial direction is 51 native LW pixels (3\farcs21). 
We subsequently extracted 1D spectra from the 2D spectra using a boxcar aperture with a height of 5 pixels (0\farcs31). Although this aperture does not capture all flux from some extended low-redshift sources, it includes the brightest central region and therefore provides higher signal-to-noise ratios for the 1D spectra, which benefits the subsequent emission-line detection. We measured line fluxes from the 2D spectra rather than the 1D spectra (see Section~\ref{sec:flux}).

\section{Sample Construction} \label{sec:sample}

In this section, we describe the main steps of our sample construction, including emission line detection, photometric redshift estimation, emission line identification and line flux measurement.

\subsection{Emission Line Detection} \label{subsec:line_detect}

Following previous works that have performed emission-line detection on NIRCam slitless spectra \citep[e.g.,][]{Wang_ASPIRE_2023,Kashino_2023,Fu_2025}, we implement the following line-detection procedure. We first remove the continuum from both the 1D and 2D spectra by applying a 1D median filter with a kernel size of 51 pixels and a central gap of 5 pixels (corresponding to the instrumental resolution of $\mathrm{FWHM}_{\mathrm{res}} \sim 29$\,\AA\ for F444W). We then smooth the continuum-subtracted spectra using a 1D/2D Gaussian kernel with a FWHM equal to the $\mathrm{FWHM}_{\mathrm{res}}$. 
To obtain robust emission-line detections, we independently perform line searches on both the 1D and 2D spectra and required consistency between the two detections.
For the 1D spectra, we identify all peaks with signal-to-noise ratio $\mathrm{SNR}>3$ in the smoothed spectra and fit a Gaussian profile to each peak to measure the line FWHM, which could be a spurious detection caused by cosmic ray. We compute the 1D line SNR by summing the flux and error within $\pm \mathrm{FWHM}$ around the fitted center. We discard lines with 1D $\mathrm{SNR}<3$ or with widths narrower than half of $\mathrm{FWHM}_{\mathrm{res}}$, as such features are likely caused by cosmic rays or other spurious artifacts.
For the 2D spectra, we use \textsc{photutils} \citep{photutils} to detect emission features, adopting $\mathrm{DETECT\_MINAREA}=5$ and $\mathrm{DETECT\_THRESH}=1.2$. We require the centroid of each detected 2D emission line to lie within 2 pixels of the spatial trace, and to match the wavelength of a 1D-detected line to within $\mathrm{FWHM}_{\mathrm{res}}$.

The combined 1D and 2D detection method described above is efficient for detecting narrow emission lines as well as broad-line AGN with a narrow component detected at $\mathrm{SNR}>3$. However, we find that the 2D detection threshold is too restrictive for broad emission features with low per-pixel SNR, such as PAH features and emission lines from low-redshift, extended star-forming galaxies. In addition, the extended wings of broad emission lines can be partially over-subtracted by the median-filtered continuum, causing AGN without a prominent narrow component to be missed. 
To mitigate these effects, we also adopt a refined detection strategy by performing detections solely on the 1D spectra. To further reduce the impact of continuum over-subtraction on broad-line wings, we apply a 1D median filter with a kernel size of 201 pixels and a central gap of 61 pixels to subtract continuum. We then detect emission lines with 1D $\mathrm{SNR}>3$. We merge the narrow- and broad-emission-line catalogs into a single master catalog.

For all detected emission features, we perform a more detailed search for potential broad components,
including intrinsically broad features such as PAH emission or from AGN, as well as those produced by morphological broadening, following the criteria similar to \citet{Jiang_2026}. In brief, we fit a linear continuum around each detected line center in the 1D spectra and subtract this continuum before line modeling. Each line is then fitted using both a single-Gaussian model with 
$\mathrm{FWHM}<12,000\ \mathrm{km}\ \mathrm{s}^{-1}$ and a double-Gaussian 
model consisting of a narrow component with 
$\mathrm{FWHM}<500\ \mathrm{km}\ \mathrm{s}^{-1}$ and a broad component 
with $500\ \mathrm{km}\ \mathrm{s}^{-1}<\mathrm{FWHM}<12,000\ 
\mathrm{km}\ \mathrm{s}^{-1}$.
The line fitting is performed using a least-squares method. The model with the smaller Bayesian Information Criterion (BIC) value is adopted as the best-fit model. Emission features whose best-fit model contains at least one component with $\mathrm{FWHM}>1000\ \mathrm{km}\ \mathrm{s}^{-1}$ are classified as broad features.
We then distinguish AGN broad lines from morphology-broadened lines and other broad emission features such as PAHs. 
%We cross-match the broad-line sources with the Chandra COSMOS Legacy Survey catalog \citep{Civano_2016} and find that 172 sources are detected in X-rays, which we identify as AGN.
PAH features are excluded from the following classification. For broad-line candidates with compactness $\mathrm{F(0\farcs7)}/\mathrm{F(0\farcs3)} < 2$, we classify them as broad-line AGN candidates. For the remaining broad-line candidates without an obvious central PSF component, we visually inspect their morphologies and remove sources showing extended structures along the dispersion direction, which are likely morphology-broadened systems rather than intrinsic broad-line AGN. The final broad-line AGN candidates are reported in our catalog.

Finally, we further remove detected lines that are highly likely spurious, such as cosmic rays or grism ghosts. Cosmic rays appear only in a single exposure, and ghost locations vary between exposures. For example, the most common ghost is the tadpole, a scattered-light artifact that can be mistaken for an emission line and appears prominently in module B. We median-stack the spectra from all exposures and perform aperture photometry at the line location on both the median- and mean-stacked spectra. If the flux difference satisfies $|\textrm{mean}-\textrm{med}| > 5\sigma$, the feature is considered highly likely to be spurious, as it does not appear consistently across exposures.
% p_med, p_all, p_mod, p_obs
Next, we compare the emission-line flux to the total F444W broadband flux. 
We convert the F444W AB magnitude to $f_\lambda$ at the filter effective wavelength and estimate the integrated broadband flux as $F_{\rm F444W}=f_\lambda\Delta\lambda_{\rm eff}$. 
If the line flux exceeds this value, the feature is considered likely contamination from another source or otherwise spurious and is removed.
%If the line flux exceeds the total F444W flux, the feature is likely contamination from another source or otherwise spurious, and we remove it.

\subsection{Photometric Redshifts and External Spectroscopic Samples} \label{subsec:photz}

We estimate photometric redshifts using the public template-fitting code \textsc{EAZY} \citep{Brammer_2008}.  We use the KRON photometry after PSF matching for \textsc{EAZY} photometric redshift estimation. We fit linear combinations of templates to our all-band photometry, allowing the redshift to vary over $0 < z < 20$.  We adopt the template set from \citet{Hainline_2023} and included a redshift prior based on the expected magnitude distribution.  We derive the photometric redshift ($z_{\mathrm{phot}}$) as the peak of the probability distribution (or $z_a$) in \textsc{EAZY} convention), and we also retain the full probability distribution $p(z)$ for uncertainty estimates and further line identification.  

We supplement our photometric redshifts with the COSMOS Spectroscopic Redshift Compilation \citep{Khostovan_2025}, which provides over 160,000 redshifts across the two–degree COSMOS field and covers a wide range of galaxy types and environments. 
% We also adopt the public data products from DESI Data Release 1 \citep{DESIDR1_2026}.
% Multiple observational programs, including the survey validation, special observations, and the main survey, cover the COSMOS field \citep{Ratajczak_2026}. Therefore, the DESI spectroscopy in the COSMOS field provides a deep and complete coverage for more than 230,000 spectroscopic redshift across the COSMOS field. 
We use these spectroscopic redshifts as an external reference set to assist the following emission line identifications in the slitless spectra.

\subsection{Emission Line Identification} \label{sec:ident}

Slitless spectroscopy efficiently captures spectra for all sources in the field without preselection, but it introduces source confusion. The dispersed traces for different sources often overlap, especially for NIRCam WFSS where the spectral resolution is high ($R\sim1600$) and the F444W band is broad. 
% A detected emission line may align with several sources along the dispersion direction, so we must decide which source is the true emitter. 
A detected emission line may be spatially consistent with multiple sources along the dispersion direction, requiring an identification of the most likely emitting source.
This is a complex and large-scale task given the large number of detected emission lines and potential counterparts. In this section, we describe how we use convolutional neural network (CNN) and LLM to assist the identification of the most likely emitting sources.

\subsubsection{Automatic Emission Line Template Cross-correlation}
\label{subsubsec:zauto}

The emission-line template that we use consists of a comprehensive set of rest-frame UV to near-IR spectral features, including hydrogen recombination lines (Balmer, Paschen, Brackett, and Pfund series),  helium lines (mostly \ion{He}{1}), nebular metal lines (e.g., \oii, \oiii, \siii), high-ionization and coronal lines associated with AGN activity (e.g., [\ion{Ne}{5}], [\ion{Si}{6}], [\ion{Fe}{10}]), molecular hydrogen and CO rovibrational transitions, and some other features such as PAH. Among these features, the primary lines used for emission-line identification include \oiii~$\lambda\lambda4959,5007$, \ha, \siii~$\lambda\lambda9069,9531$, \ion{He}{1}~$\lambda10830$, \pab, \paa, \brb, PAH, and \bra, while the remaining lines serve as auxiliary constraints.
This template enables emission-line identification across a wide redshift range.

For each source with emission lines detected in both the 1D and 2D NIRCam F444W slitless spectra, we first run the automatic algorithm described in Lin et al. (in prep.), following the same procedure as \citet{Sun_2025_SAPPHIRES}. 
In brief, the algorithm performs a two-step cross-correlation between the detected lines and a set of emission-line templates. The first step matches line positions, and the second refines the redshift by matching line-flux ratios and combining the result with the photometric redshift. This algorithm works well when the photometric redshift is reliable or when multiple emission lines fall within the F444W band. 
However, most spectra contain only a single detectable line, the photometric redshift sometimes shows multiple probability peaks, and some detected lines originate from nearby sources aligned along the dispersion direction, which cannot be easily removed with only Grism~R spectra. These limitations make additional inspection essential.

\begin{figure*}[t]
\centering
\includegraphics[width=\textwidth]{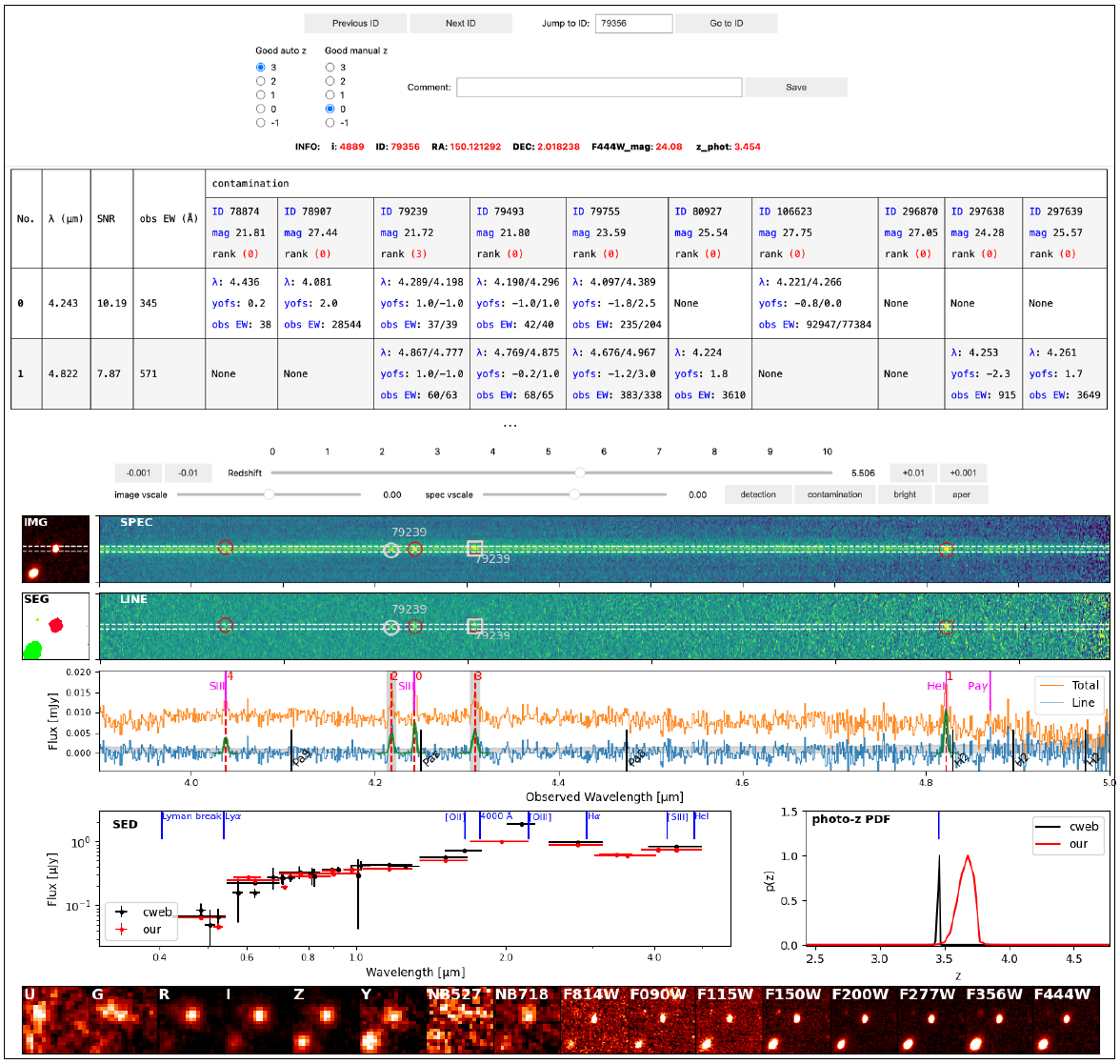}
\caption{Visual inspection interface for an example galaxy at $z = 3.451$. 
The upper panel shows the widgets and the contamination table (lines 2-4 are omitted for display purpose).
The lower panel displays, from top to bottom and left to right: 
(1) a cutout of the direct image in the F444W band, 
(2) the combined 2D spectrum with detected emission lines marked by red circles and contamination indicated by gray circles or boxes, 
(3) the cutout of the segmentation image, 
(4) continuum-subtracted line spectra, 
(5) the combined 1D spectrum with detected emission lines marked by red dashed lines and contamination shown as gray shaded regions, 
(6) spectral energy distribution (SED), 
(7) the PDF of the photometric redshift, and 
(8) cutouts of the direct images in multiple bands. 
The theoretical positions of the line features at this redshift are marked with short vertical lines in panels (5)–(7), and update automatically when the user adjusts the redshift control bar. \label{fig:VI_tool}
}
\end{figure*}

\subsubsection{First-round Visual Inspection}

%We first visually examine 10\% of the full sample.
We first visually inspect all sources with emission-line $\mathrm{SNR}>5$ in half of the survey area.
The visual inspection interface is shown in Figure~\ref{fig:VI_tool}. 
For each source, we use spectral tracing functions to identify all other sources whose spectral traces intersect the line, including the genuine emitting source and possible contaminating sources. All of them are listed in the above table. In this way, we can carefully compare and determine which source produces each line, and assign redshifts to all objects in the group. 
For each potential source, the table provides the corresponding wavelength of the line, the offset from the trace center of the target source in the perpendicular direction (y-offset), and the observed equivalent width (obs EW), computed from the line flux and the F444W broadband flux after subtracting the line contribution.
Users can directly check any source using the ID to compare and identify which source the line belongs to.

The identification is based on all information: 
\begin{itemize}
\item The morphological similarity between the F444W image and the spatial profile of the emission line, which is particularly important for extended low-redshift sources.
\item The EWs of the emission lines, which help distinguish high-redshift and low-redshift line identifications, given the prior that high-redshift sources (e.g., \ha\ and \oiii\ emitters) are more likely to exhibit large observed EWs.
\item The full multi-band SED and imaging data, including both the COSMOS2025 photometry and our own measurements.
\item The photometric redshift and its probability distribution function (PDF), which are critical when multiple redshift solutions arise from degeneracies such as the Lyman break versus Balmer break.
\item Ground-based spectroscopic redshifts from \citet{Khostovan_2025}, when available. These measurements are useful for validation, although some low-confidence redshifts can be unreliable.
\item Spectra from all individual exposures, together with their median-stacked spectra (although not shown in Figure~\ref{fig:VI_tool}), which help identify and remove remaining spurious detections such as cosmic rays, hot pixels, and scattered light artifacts.
\end{itemize}
%morphological similarity between F444W image and spatial profile of the emission-line, which is important for low-redshift extended sources; 
%the EWs of the emission lines, which help distinguish the high-redshift and low-redshift emission lines (with the prior that high-redshift sources, e.g., \ha and \oiii\ emitters, are more likely to have large EWs); the full multi-band SED (both COSMOS2025 and our photometry measurements are shown) and imaging; 
%the photometric redshift and its probability distribution function (PDF), which is critical when multiple peaks arise from degeneracies such as Lyman-break versus Balmer-break; 
%ground-based spectroscopic redshifts from \citet{Khostovan_2025} and DESI when available, which is useful but sometimes unreliable with low confidence level; 
%spectra from all individual exposures, as well as their median-stacked spectra, are also available for inspection (although not shown in Figure~\ref{fig:VI_tool}), which help identify and remove remaining spurious detections (e.g., cosmos-ray, hot pixels, scatter light, etc.).
The initial redshift is taken from the results of the automatic algorithm, and users may adjust the redshift for verification.
Once a line is assigned to a source, it is marked in the 2D spectrum by a gray circle (for location in module A) or a box (module B), and in the 1D spectra of contamination sources by a gray shaded region, to prevent multiple assignments.
The first round identifies 4,422 line emitters. %\textcolor{red}{the number a useful information in CNN training)}

\begin{figure}[!t]
\centering
\includegraphics[width=0.45\textwidth]{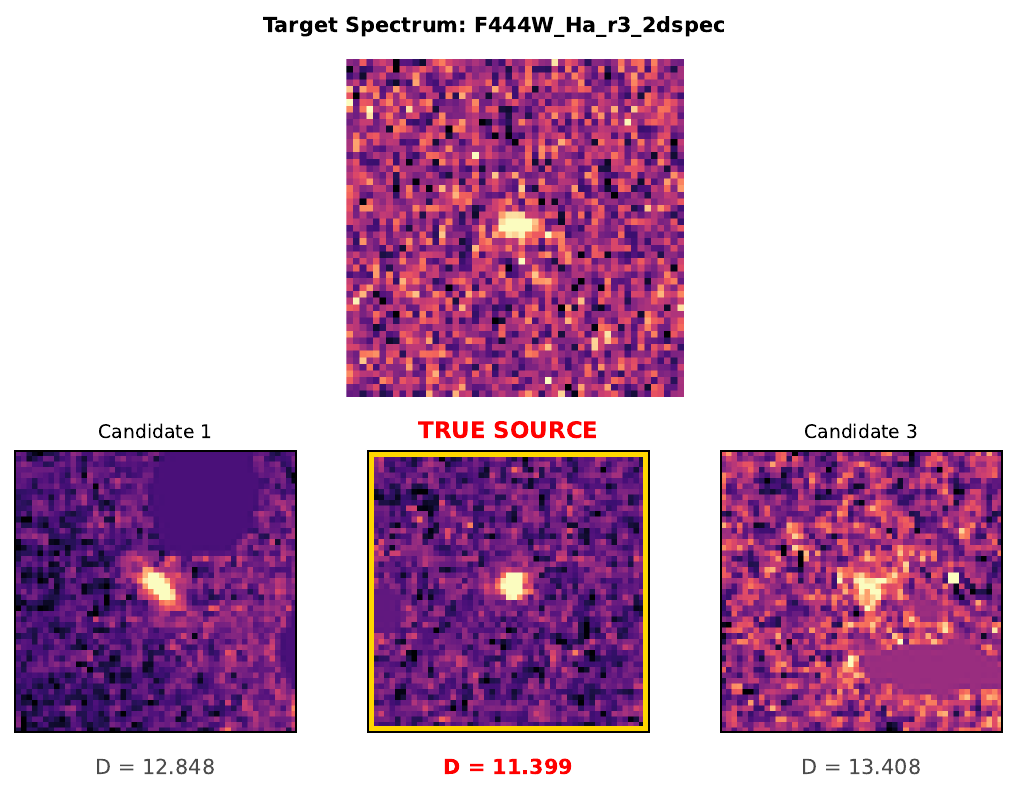}
\caption{Demonstration of the Siamese morphology similarity distance. The top panel shows the 2D cutout of the 2D spectrum of an example target emission line. The subsequent rows display the images of the potential source candidates. The second candidate highlighted with a yellow border represents the real source of the emission line, while the others are contaminations. The value below each panel indicates the Euclidean distance between the images and the target 2D spectrum. The real source has the lowest Siamese distance.}
\label{fig:CNN}
\end{figure}

\subsubsection{Siamese Network: Morphology Similarity between 2D Spectrum and Image}
\label{subsubsec:CNN}
Using the human-confirmed subsample, we train a Siamese network architecture \citep{bromley1993signature} with a convolutional neural network (CNN) backbone to quantify the similarity between each galaxy image and the 2D emission-line profile. 
The training set from the visually inspected subset provides reliable labels, including 4,422 positive sources and 12,336 negative sources.
For each source, we extract cutouts from the imaging and from the spectra around the emission-line region. In the imaging cutouts, we mask neighboring sources. The spectral cutouts are taken from each exposure and then stacked. 
For Grism~R data, the dispersion directions of module~A and module~B are opposite, so the module~B spectral cutouts are flipped left–right to match the morphology in module~A before stacking. In addition, we account for the telescope position angle in each exposure and rotate the cutouts to a common reference frame.

Our network is based on a pretrained EfficientNet-B0 backbone, implemented using the \texttt{timm} (PyTorch Image Models) library\footnote{\url{https://github.com/huggingface/pytorch-image-models}.}. 
We further train the network to distinguish matched (same-source) and unmatched (contamination) 2D emission-line and direct-image pairs, enabling a quantitative assessment of morphological consistency between galaxy images and their emission-line structures. 
Specifically, the Siamese network is optimized such that the feature embeddings of positive pairs are pulled closer together in the latent space, while those of negative pairs are pushed farther apart. This is achieved through a contrastive loss function, which enforces a small distance for matched pairs and a large distance (up to a margin) for unmatched pairs.
Figure \ref{fig:CNN} shows an example of the comparison of contamination and real sources, with the real source showing the lowest similarity distance to the emission line. 
Our final Siamese network achieves an accuracy of $\sim 80\%$ in distinguishing genuine sources from contaminants. Most misclassifications arise in cases where contaminants exhibit morphologies closely resembling those of real sources; such contaminants can be further distinguished using additional criteria like EWs, photometric redshift, and y-offset.

\subsubsection{LLM-assisted Emission Line Identification}

We then use an LLM agent to emulate the human emission–line identification process. 
The adoption of LLMs in this context follows the recent trend of integrating LLMs into diverse astronomical tasks, where they have demonstrated significant potential in processing complex datasets and automating decision-making workflows \citep[e.g.,][]{Sun2025,Ting2025}.
Specifically, we use the \textit{doubao-seed-1-6-251015} model, developed by ByteDance\footnote{\url{https://www.volcengine.com/product/ark}}. 
%Our selection of this model followed a comprehensive performance benchmark against other leading architectures, including GLM and Qwen, which revealed that the Doubao model offers the best performance in our task while maintaining significant cost-effectiveness. 
We select this model after benchmarking against several other leading large language models, including the ChatGLM, Gemini, and Qwen model families. The Doubao model achieves one of the best overall performances for our task while remaining significantly more cost-effective.
In our pipeline, all information relied upon during the human visual inspection process is quantitatively encoded into a structured JSON format (see the Input Format section of the prompt
%in Appendix \ref{appendix:LLM_prompt} 
for details), including the morphological consistency quantified by Siamese distance in Section \ref{subsubsec:CNN}.
These quantified attributes allow the LLM to learn decision patterns similar to those used in human inspection. 

The LLM is guided by a sophisticated prompt designed to enforce physical consistency and disentangle potential contamination. % (Add online source later). %(Appendix \ref{appendix:LLM_prompt}). 
Briefly, the model is instructed to act as an expert assistant, synthesizing evidence from multiple aspects. It compares the $z_{\mathrm{auto}}$ (Section \ref{subsubsec:zauto}) against photometric constraints ($z_{\mathrm{phot}}$), evaluates whether a detected line's observed EW is physically plausible or supported by broadband SED flux excesses, and utilizes the relative Siamese distance to weigh morphological probabilities, etc. 
The prompt implements a mandatory line-uniqueness and assignment check, requiring the model to assign at least one source (and exactly one source when the rank is 3) to every real emission line detected. 
This ensures that overlapping spectra from nearby sources are correctly attributed or flagged as contamination.

The integration of this LLM-based agent has significantly accelerated our data processing workflow and reduced human workload, achieving a throughput of approximately 850 source groups per day, with each group encompassing both primary targets and their associated contaminants. 
%The operational cost is remarkably low, averaging approximately 0.1 CNY ($\sim$0.014 USD) per source group.
The emission line sample with redshift determined by the LLM is further visually inspected by multiple authors using the visual inspection interface. 
The initial redshift is taken from the LLM result, which is correct in approximately 90\% of cases (see Section~\ref{sec:z_compare}), while maintaining the rigorous standards of human oversight.

\subsection{Line Flux Measurement} \label{sec:flux}  

We use \textsc{photutils} \citep{photutils} on the 2D spectra to perform source detection and measure the fluxes of emission lines.
While the median-filter--based continuum subtraction is sufficient for emission-line detection, it can lead to a flux loss of up to $\sim10\%$ due to oversubtraction in the vicinity of the line.
To mitigate this effect, we adopt a more precise procedure to model and subtract the local continuum prior to flux measurements.

For broad emission lines, the SNR per pixel is low, while the total line flux and integrated SNR remain high.
Prior to flux measurements, the 2D spectra are rebinned along the dispersion direction using an adaptive bin size.
For each emission line, the number of spectral pixels combined into a single bin is defined as
$N_{\mathrm{bin}} = \mathrm{FWHM}/80~\mathrm{\AA}$.
The rebinned pixel values are summed rather than averaged, ensuring that the total flux measured in the 2D spectra is conserved.
The associated variance maps are propagated through the same rebinning procedure.
%Correspondingly, the variance maps are rebinned assuming uncorrelated noise between adjacent spectral pixels.

We then fit a cubic polynomial to the continuum in each spatial row of the 2D spectra, using regions adjacent to the emission line ($\pm 60$ pixels, corresponding to $\pm (600 \times N_{\mathrm{bin}})$~\AA).
Source detection is subsequently performed on the continuum-subtracted spectra to generate a segmentation map, which is used to mask pixels associated with the emission line.
The continuum fitting is then repeated with the masked regions excluded.
This procedure is iterated until convergence, after which the emission-line flux is measured from the final continuum-subtracted spectra.
To account for both the spatial distribution and the morphology broadening effect of the slitless spectra, we obtain Kron-aperture photometry \citep{Kron_1980} with a Kron factor of 3, which is large enough to include nearly all flux while maintaining a high SNR.  

%The flux measurements of low-SNR emission lines are often artificially boosted due to random noise. If not corrected, this can overestimate the volume density of more luminous sources, which is also known as the Eddington bias \citep{eddington1913}. To quantify this effect, we use a Monte Carlo (MC) method and compare injected and output line flux for a range of line SNRs from 3 to 50. For each input SNR, we apply a correction to our sources using the median value of the output-to-input line flux ratios. The correction factor for the Eddington bias is found to be xxx for sources with SNR = 3, and xxx for those with SNR = 10. Approximately half of the sources in our sample have SNR values between 5 and 10, and are therefore subject to moderate levels of the Eddington bias.

\begin{figure}[!t]
\centering
\includegraphics[width=0.48\textwidth]{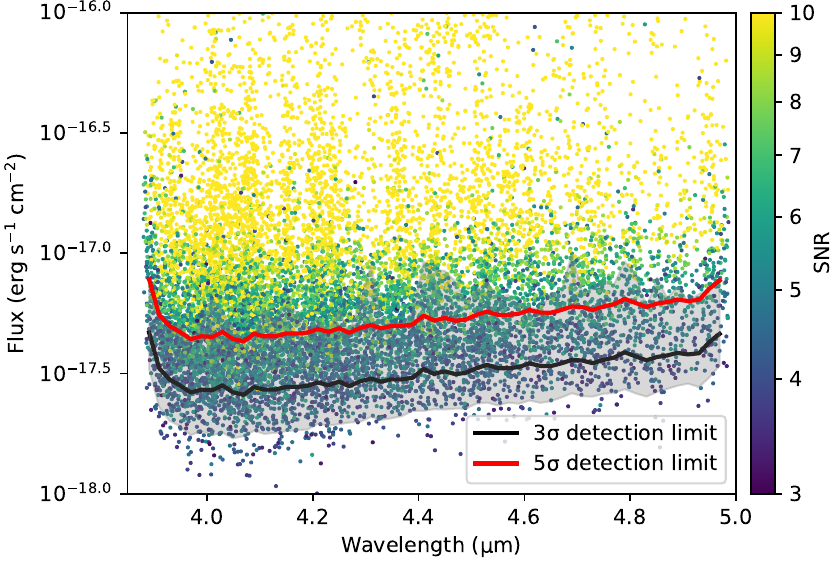}
\caption{Distribution of observed wavelength and flux of detected emission lines in the survey. The scatter points represent individual detected lines, colored by their SNR. The solid black (red) line shows the median 3$\sigma$ (5$\sigma$) detection limit, and the shaded region corresponds to the 1$\sigma$ uncertainty range of the 3$\sigma$ detection limit.   \label{fig:detect_lim}}
\end{figure}

\subsection{SED Modeling} \label{sec:sed}

We perform SED modeling with \textsc{bagpipes} \citep{Carnall_2018} for the emission-line galaxies. 
For each source, the redshift is fixed to the spectroscopic redshift. 
We use the available multi-band photometry from Subaru/HSC, HST/ACS, and JWST/NIRCam. 
For sources with robust detections of hydrogen recombination lines, we also include the observed line fluxes in the fit to better constrain the recent star formation activity. 
The fitted lines include H$\alpha$, H$\beta$, Pa$\alpha$, Pa$\beta$, Br$\alpha$, and Br$\beta$, depending on the available line detections for each galaxy.

We use the \citet{Bruzual_2003} stellar population synthesis models and adopt the \citet{Kroupa_2001} initial mass function. 
We adopt the non-parametric continuity star formation history implemented in \textsc{bagpipes}. 
The star formation history is divided into age bins with edges at 0, 3, 10, and 100 Myr, followed by three broader bins between 100 Myr and the age of the Universe at the source redshift \citep{Leja_2019}. 
The ratios of the star formation rates between adjacent time bins are allowed to vary over a broad range, with a Student-$t$ prior imposed on each \texttt{dsfr} parameter. 
The stellar metallicity is allowed to vary from 0.001 to 2.5 $Z_\odot$, with a logarithmic prior.
Dust attenuation is modeled using the Salim attenuation law \citep{Salim_2018}. 
We allow the $V$-band attenuation to vary over $A_V=0$--4 mag. 
The extra attenuation in birth clouds is controlled by the parameter $\eta$, which is allowed to vary between 1 and 3. 
We also allow the slope modification of the attenuation curve to vary over $\delta=-0.3$--0.3, and the strength of the UV bump to vary over $B=0$--5. 
Nebular emission is included, with the ionization parameter allowed to vary over $\log U=-4$ to 0.

For each fitted source, we derive posterior distributions for the fitted model parameters and a set of physical quantities, including stellar mass, formed stellar mass, star formation rate, specific star formation rate, SFR averaged over the last 10 Myr, mass-weighted age, formation time, quenching time, and mass-weighted metallicity. 
We also compute rest-frame colors, UV-continuum properties, ionizing photon production efficiencies, and model emission-line fluxes and equivalent widths. 
In the catalog, we report the 16th, 50th, and 84th percentiles of these posterior distributions, together with the photometric $\chi^2$ and the number of fitted bands. %Bayesian evidence of each fit.

\section{Results} \label{sec:result}

\begin{figure}[!t]
\centering
\includegraphics[width=0.49\textwidth]{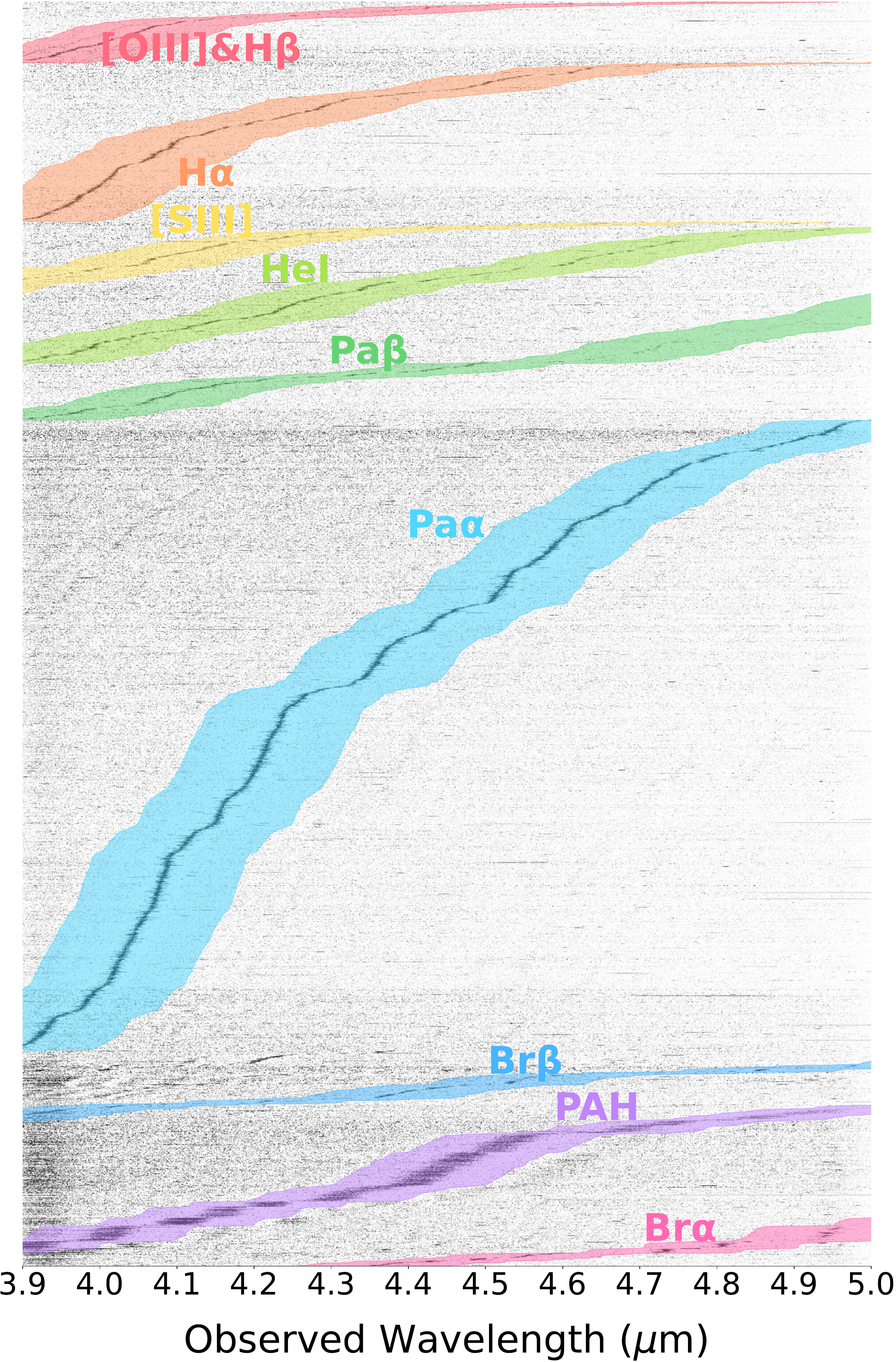}
\caption{ Spectral sample with the highest confidence level (\texttt{rank = 3}) and at least one emission line with $\mathrm{SNR} > 7$. The most prominent emission lines at each redshift are labeled.   \label{fig:all_spec}}
\end{figure}

\subsection{Line-emitter Catalog} \label{subsec:line_cat}
We present our spectroscopic redshift catalog, which includes 12,275 emission-line sources. 
The catalog format is summarized in Table~\ref{tab:sample}. 
For each source, we provide its COSMOS-Web ID, the corresponding ID in the photometric catalog used in this work, sky coordinates, spectroscopic redshift, redshift quality rank, identified emission-line species, observed line center, line flux, and line-flux uncertainty. 
We also include ancillary information, including the broad-line AGN flag, photometric redshift and its confidence interval, F444W magnitude, possible contamination from nearby sources, and physical properties derived from SED modeling, such as stellar mass, recent 10 Myr SFR, color excess, and stellar population metallicity. 
This catalog therefore provides both a spectroscopic redshift sample and a homogeneous set of emission-line and galaxy-property measurements for statistical studies of galaxies and AGN.

All sources in the catalog are assigned a quality rank from 1 to 3 based on the reliability of their redshift identification. 
Sources with \texttt{rank = 3} have the highest confidence, either exhibiting more than one spectral feature or showing a single line with a spectroscopic redshift consistent with the photometric redshift, and no highly probable contamination. 
Sources with \texttt{rank = 2} have reasonably high confidence, typically characterized by a single detected emission line that is close to the photometric redshift and/or the presence of possible contamination. 
Sources with \texttt{rank = 1} represent tentative but plausible identifications, for which the preferred redshift solution is less strongly constrained by the photometric redshift or potential contamination cannot be confidently excluded. 
In some cases, it is difficult to determine which source an emission line belongs to. 
In such situations, both sources are assigned \texttt{rank = 1} or \texttt{rank = 2}, and they are cross-referenced by listing each other in the \texttt{contm\_ID} column. 
The \texttt{rank = 3} sources form the most secure subset of the catalog, while the lower-rank sources are retained to provide a more complete record of possible line detections and ambiguous cases.

About four thousand 2D spectra with the highest confidence (\texttt{rank = 3}) and at least one emission line detected with ${\mathrm{SNR}} > 7$ are shown in Figure~\ref{fig:all_spec}, ordered from top to bottom by decreasing redshift. 
These spectra illustrate the wide range of emission-line features covered by the survey, from low-redshift hydrogen recombination lines in the Paschen and Brackett series to rest-frame optical lines such as \ha, \hb and \oiii\ $\lambda\lambda4959,5007$ at higher redshifts. 
The broad wavelength coverage of the NIRCam grism data allows different transitions to be observed over complementary redshift intervals, enabling a nearly continuous census of line-emitting galaxies from the local Universe to the epoch of reionization.

\begin{figure}[!t]
\centering
\includegraphics[width=0.49\textwidth]{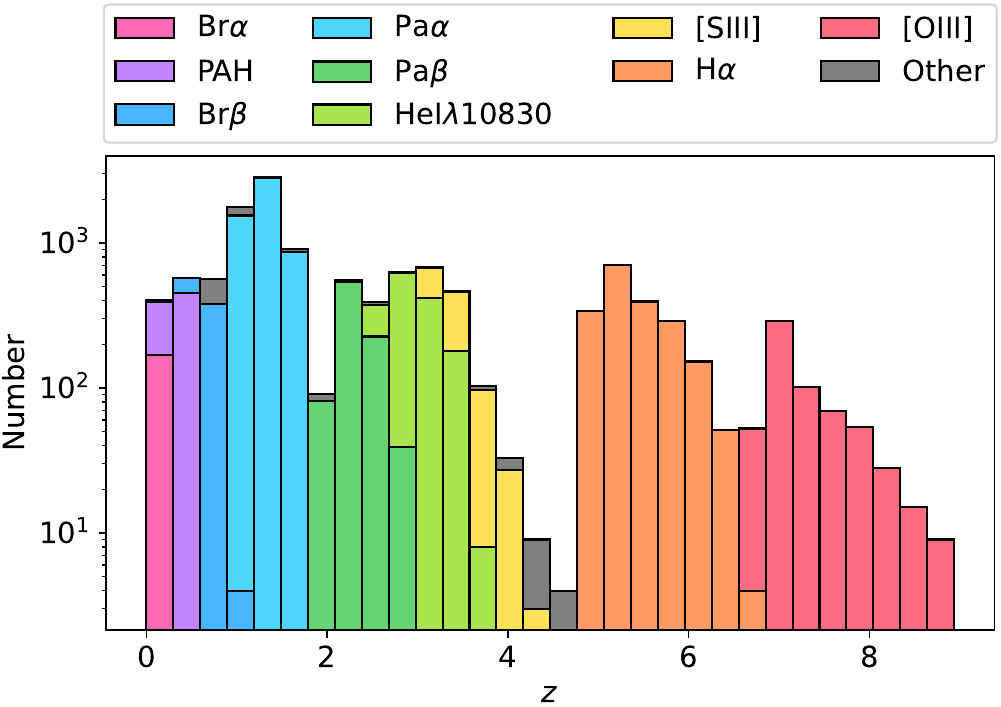}
\caption{ Redshift distribution of the line-emitter sample. Bars in different colors represent different primary emission lines in the spectra.   \label{fig:z_hist}}
\end{figure}

The redshift distribution of the full sample is presented in Figure~\ref{fig:z_hist}. 
Most sources are confirmed by one or more primary emission lines, ranging from \bra\ to \oiii\ $\lambda\lambda4959,5007$. 
The relative numbers of emitters selected through different lines are set by a combination of intrinsic line strengths, survey sensitivity, wavelength coverage, dust attenuation, and the cosmic evolution of the galaxy population. 
Among these, \paa\ emitters constitute the largest fraction, consistent with the elevated star formation rate density at cosmic noon. 
At the same time, the broad-line flag and the presence of high-ionization emission lines make the catalog useful for identifying AGN and composite systems.

The homogeneous selection and ranking scheme make the catalog a versatile resource for future studies of star formation, black hole growth, and the connection between galaxies and their environments across cosmic time.
In the Section~\ref{subsec:sci}, we outline several science applications enabled by the catalog.

\begin{deluxetable}{ll}
\tablecaption{Columns in the emission line catalog \label{tab:sample}}
%\tablewidth{0pt}
\tablehead{
\colhead{Label} & \colhead{Description}
}
\startdata
\texttt{ID} & COSMOS-Web ID \\
\texttt{ID\_phot} & ID in photometric catalog this work \\
\texttt{RA} & Right Ascension (J2000) in degrees \\
\texttt{DEC} & Declination (J2000) in degrees \\
\texttt{z} & Spectroscopic redshift \\
\texttt{rank} & Ranking of the redshift \\
\texttt{line\_name} & Names of the detected emission lines \\
\texttt{line\_center} & Wavelengths of the line centers in \AA \\
\texttt{line\_flux} & Line fluxes in $10^{-17}$ erg s$^{-1}$ cm$^{-2}$ \\
\texttt{line\_flux\_err} & 1$\sigma$ uncertainties of the line fluxes \\
\texttt{flag\_broad\_line} & Broad-line AGN flag \\
\texttt{z\_phot} & Photometric redshift of the source \\
\texttt{z\_phot\_16} & 16th percentile of photometric redshift \\
\texttt{z\_phot\_84} & 84th percentile of photometric redshift \\
%\texttt{F444W\_mag} & Magnitude in NIRCam F444W band \\
%\texttt{F444W\_mag\_err} & Uncertainty of F444W magnitude \\
\texttt{contm\_ID} & IDs of possible contamination sources \\
\texttt{log\_mass} & Stellar mass in \msun \\
\texttt{log\_SFR\_10Myr} & Recent 10~Myr SFR in \smpy \\
\texttt{log\_SFR\_100Myr} & Recent 100~Myr SFR in \smpy \\
\texttt{A\_V} & $V$-band attenuation, $A_V$ \\
\texttt{log\_Z\_stellar} & Stellar metallicity, $\log(Z_\star/Z_\odot)$ \\
\texttt{chisq\_phot} & $\chi^2$ of the best-fit SED model \\
\texttt{n\_bands} & Number of bands used in SED fit \\
%\texttt{log\_mass} & Stellar mass in \msun \\
%\texttt{log\_mass\_err} & Stellar mass uncertainty \\
%\texttt{log\_SFR\_10Myr} & Recent (10~Myr) SFR in \texttt{smpy}\\
%\texttt{log\_SFR\_10Myr\_err} & SFR uncertainty \\
%\texttt{EBV} &  $E(B-V)$ \\
%\texttt{EBV\_err} &  $E(B-V)$ uncertainty \\
%\texttt{log\_Z\_stellar} & stellar population metallicity \\
%\texttt{log\_Z\_stellar\_err} & metallicity uncertainty \\
\enddata
\tablecomments{
For SED-derived quantities, suffixes \texttt{\_16}, \texttt{\_50}, and \texttt{\_84} denote the 16th, 50th, and 84th percentiles of the posterior distribution, respectively.
}
\end{deluxetable}

\subsection{Composite Spectra} \label{sec:stack_spec}
We construct the composite spectra by median stacking the normalized 2D spectra of our sample. Compared with the mean-stacking method, the median-stacking method avoids the spectral features being dominated by a few outliers in the sample. In addition, the 2D-stacking method allows us to inspect the composite 2D spectrum to verify the reliability of emission lines.

As shown in Figure~\ref{fig:z_hist}, our sample is concentrated in several redshift intervals due to a few prominent emission lines. 
%We bin the sample by their prominent emission lines: 
We therefore define several redshift intervals based on the prominent emission lines covered in each range: (1) \bra, PAH, and \brb, which covers $z=0-0.83$; (2) \paa, which covers $z=1.10-1.56$; (3) \pab, \ion{He}{1} $\lambda10830$, and [\ion{S}{3}] $\lambda\lambda9069,9531$, which covers $z=2.07-3.99$; (4) \ha, which covers $z=5.00-6.31$; (5) \oiii $\lambda\lambda4959,5007$ and \hb\, which covers $z=6.87-8.58$. 

In each redshift interval, for each observed spectrum, we use cubic spline interpolation to shift the spectrum to the rest frame on a common rest-frame wavelength grid with a wavelength interval of $10\,\mathrm{\AA}/(1+\bar{z})$, where $\bar{z}$ is the mean redshift of this subsample. 
We normalize each spectrum based on the galaxy’s F444W photometry. We subtract the continuum before stacking for the $z>2$ bins.
An associated variance image for each observed spectrum is generated based on the error extension and is also shifted to the common wavelength grid. The error spectrum for the final stacked spectrum is constructed by propagating the variance spectrum of each object.

While the flux ratios of emission lines between different composite spectra are not directly comparable, owing to the different redshifts of the samples and the broadband-based normalization applied to each spectrum, the relative flux ratios within each composite spectrum can still provide useful information. We measure emission-line fluxes by fitting the 1D spectra with Gaussian profiles, and by using Drude profiles to model the PAH and aliphatic features. For the $z<2$ bins, we construct a median continuum after masking the emission-line and CO absorption regions, and subtract it from the spectra. For isolated emission lines, we fit each line with a single Gaussian component and fix the central wavelength.
For blended lines separated by less than three times the FWHM, such as \ha\ and [\ion{N}{2}]$\lambda\lambda6549,6585$, \paa\ and \ion{He}{1}$\lambda18685$, as well as the PAH and aliphatic features, we perform simultaneous fits using multiple Gaussian (or Drude) components with fixed central wavelengths. We report the fluxes of all lines relative to the main line in each redshift bin (\bra\, \paa, \ion{He}{1}$\lambda10830$, \ha, and \oiii $\lambda5007$) in Table~\ref{tab:line_stack}.

\begin{figure*}[!t]
\centering
\includegraphics[width=\textwidth]{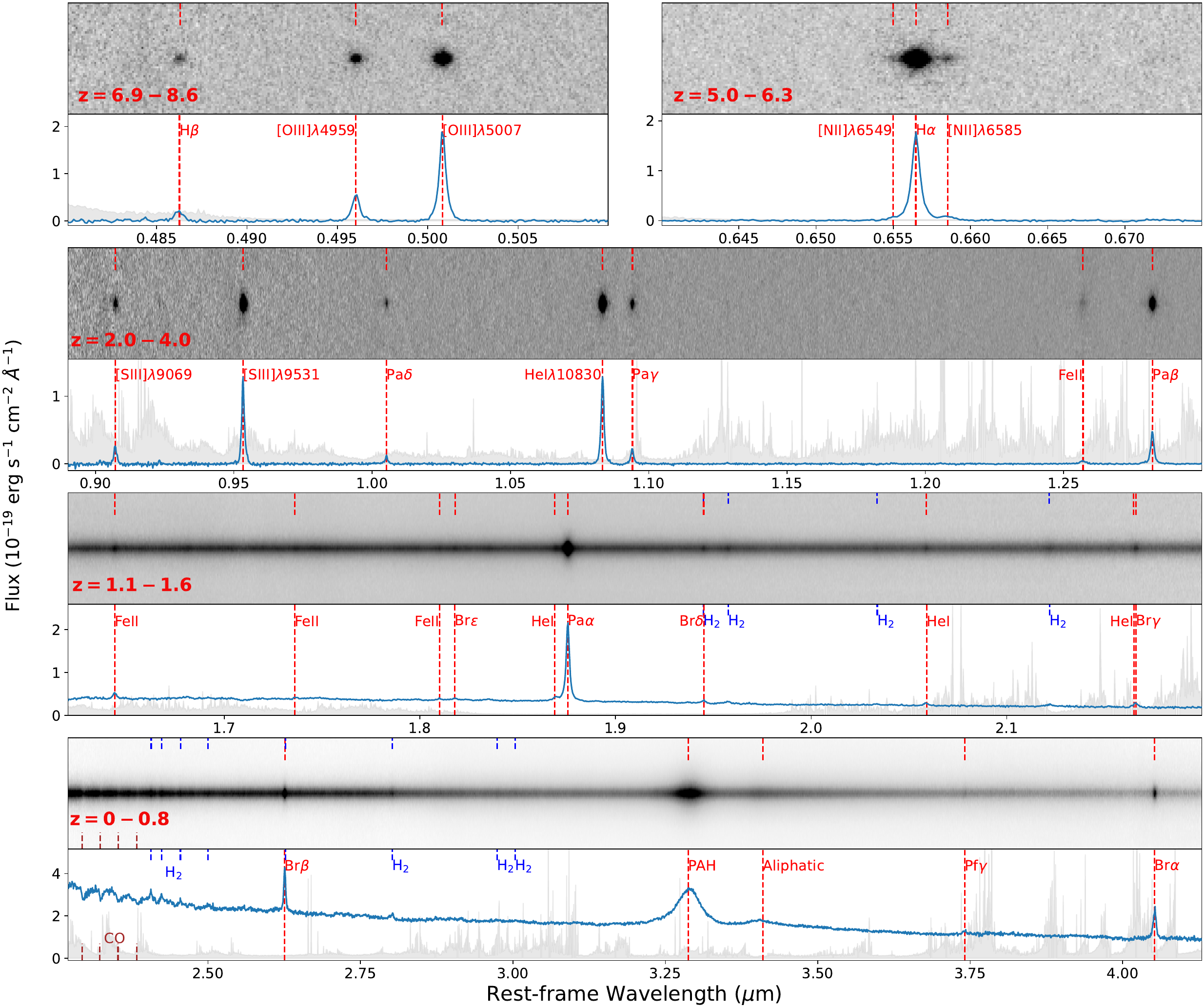}
\caption{ Median composite spectra in different redshift bins, showing both the stacked 2D spectra (top panels in each row) and the corresponding 1D spectra (bottom panels). Prominent emission and absorption features are labeled in each redshift interval, including hydrogen recombination lines, forbidden metal lines, molecular hydrogen transitions, and PAH features. The blue curves show the median stacked spectra, while the gray shaded regions indicate the scatter among individual spectra. Vertical dashed lines mark the expected wavelengths of identified spectral features.  \label{fig:stack_spec}}
\end{figure*}

\begin{deluxetable}{lll}
\tablewidth{\columnwidth}
\tablecaption{Line Flux Ratio in the composite spectra \label{tab:line_stack}}
\tablehead{
\colhead{Line} & \colhead{$\lambda$ ($\mu$m)} & \colhead{$100\times F(\lambda)/F(\mathrm{main})$}
}
\startdata
\multicolumn{3}{c}{$z=6.9$--$8.6$} \\
\hline
\oiii\ $\lambda5007$ & $0.5008$ & $100.00$ \\
\oiii\ $\lambda4959$ & $0.4960$ & $32.12$ \\
\hb & $0.4863$ & $13.70$ \\
\hline
\multicolumn{3}{c}{$z=5.0$--$6.3$} \\
\hline
\nii\ $\lambda6585$ & $0.6585$ & $10.09$ \\
\ha & $0.6565$ & $100.00$ \\
\nii\ $\lambda6549$ & $0.6550$ & $10.20$ \\
\hline
\multicolumn{3}{c}{$z=2.0$--$4.0$} \\
\hline
\pab & $1.2822$ & $42.21$ \\
\ion{Fe}{2} $\lambda12570$ & $1.2570$ & $6.57$ \\
Pa$\gamma$ & $1.0941$ & $17.09$ \\
\ion{He}{1} $\lambda10830$ & $1.0833$ & $100.00$ \\
Pa$\delta$ & $1.0052$ & $8.02$ \\
\siii\ $\lambda9531$ & $0.9533$ & $89.86$ \\
\siii\ $\lambda9069$ & $0.9071$ & $19.95$ \\
\hline
\multicolumn{3}{c}{$z=1.1$--$1.6$} \\
\hline
Br$\gamma$$+$\ion{He}{1} $\lambda21650$ & $2.1661$ & $6.89$ \\
$\mathrm{H}_2$ 1-0 S(1) & $2.1218$ & $4.40$ \\
\ion{He}{1} $\lambda20590$ & $2.0590$ & $4.47$ \\
$\mathrm{H}_2$ 1-0 S(2) & $2.0338$ & $1.65$ \\
$\mathrm{H}_2$ 1-0 S(3) & $1.9576$ & $5.50$ \\
Br$\delta$$+$$\mathrm{H}_2$ 2-1 S(5) & $1.9451$ & $3.47$ \\
\paa & $1.8756$ & $100.00$ \\
\ion{He}{1} $\lambda18685$ & $1.8689$ & $9.12$ \\
Br$\epsilon$ & $1.8179$ & $1.73$ \\
\ion{Fe}{2} $\lambda18100$ & $1.8100$ & $1.20$ \\
\ion{Fe}{2} $\lambda17360$ & $1.7360$ & $1.20$ \\
\ion{Fe}{2} $\lambda16440$ & $1.6440$ & $6.66$ \\
\hline
\multicolumn{3}{c}{$z=0$--$0.8$} \\
\hline
\bra & $4.0523$ & $100.00$ \\
Pf$\gamma$ & $3.7410$ & $12.16$ \\
Aliphatic & $3.4100$ & $971.22$ \\
PAH & $3.2880$ & $1039.58$ \\
$\mathrm{H}_2$ 1-0 O(4) & $3.0039$ & $7.22$ \\
$\mathrm{H}_2$ 2-1 O(3) & $2.9741$ & $6.92$ \\
$\mathrm{H}_2$ 1-0 O(3) & $2.8025$ & $9.73$ \\
\brb$+$$\mathrm{H}_2$ 1-0 O(2) & $2.6259$ & $110.10$ \\
$\mathrm{H}_2$ 1-0 Q(7) & $2.5001$ & $15.03$ \\
$\mathrm{H}_2$ 1-0 Q(5) & $2.4548$ & $14.76$ \\
$\mathrm{H}_2$ 1-0 Q(3) & $2.4237$ & $16.69$ \\
$\mathrm{H}_2$ 1-0 Q(1) & $2.4066$ & $22.49$ \\
\enddata
\end{deluxetable}

\subsection{Science Enabled by the Emission-Line Catalog} \label{subsec:sci}

The emission-line catalog presented in this work opens a broad range of scientific applications, enabled by its homogeneous line identification, quality ranking, and joint measurements of spectroscopic redshifts, line fluxes, and galaxy physical properties.

A key strength of the catalog is the simultaneous detection of multiple hydrogen recombination lines, including the Balmer, Paschen, and Brackett series, from star-forming galaxies over a broad redshift interval spanning $0 \lesssim z \lesssim 9$. 
By identifying and spectroscopically confirming emission-line galaxies with a uniform selection methodology, we construct a unified sample of star-forming galaxies that probes nearly the full history of cosmic star formation. 
This sample enables a systematic study of the star-forming main sequence across cosmic time. 
We will also measure emission-line luminosity functions in successive redshift windows and derive the cosmic star formation rate density in a fully self-consistent framework. 
The use of multiple recombination lines allows us to trace star formation over a wide dynamic range in dust attenuation and intrinsic luminosity, providing a more complete census of star-forming galaxies than is achievable with any single indicator alone.

Beyond the star-forming galaxy population, this catalog also contains a diverse sample of broad-line sources. 
These include radio AGN, blue quasars, red quasars, high-redshift AGN, and little red dots. 
Their physical properties, including stellar populations, dust attenuation, AGN contribution, and black hole accretion properties, require dedicated modeling beyond the scope of this catalog paper. 
They will therefore be analyzed separately with tailored SED fitting and spectral decomposition in a series of forthcoming studies (\citealt{2026NatAs.tmp..118F}, Z. Zhang et al. in prep.; D. Jiang et al. in prep.).

The continuous sky coverage of the emission-line samples also provides an ideal data set for studying the connection between galaxy growth and environment. 
Because line emitters are selected over contiguous areas and over well-defined redshift intervals, the catalog can be used to identify galaxy overdensities, trace large-scale structures, and quantify how star formation and nuclear activity depend on local environment. 
The environmental properties of the line-emitting galaxies, including their relation to large-scale structures, will be presented in future work (D. Jiang et al. in prep.). 
On smaller physical scales, the same catalog enables systematic searches for close galaxy pairs and interacting systems. 
Such samples will provide a direct view of the connection between galaxy interactions, star formation, and AGN activity across cosmic time.
%Such samples will provide a direct view of merger-driven star formation and AGN triggering across cosmic time, and will be explored in Y. Wang et al. in prep.

Finally, the large number of spectroscopically confirmed emitters enables high-SNR composite spectra in different galaxy populations and redshift bins. 
As shown in Figure~\ref{fig:stack_spec}, the stacked spectra reveal a rich set of weak emission and absorption features, including the H$_2$ series, CO absorption bands, and \ion{Fe}{2} lines. 
These features provide access to physical processes that are difficult to study in individual faint galaxies. 
In particular, the detected H$_2$ rovibrational emission lines offer a valuable probe of the warm molecular gas content and excitation mechanisms in these systems, which will be investigated in J. Chen et al. in prep.

\subsection{Data release} \label{subsec:data_release}

The main data products presented in this work will be publicly released upon publication. 
These products include the reduced JWST imaging mosaics, the multi-band photometric catalog, the extracted NIRCam grism spectra, and the final line-emitter catalog. 

The released photometric catalog will include the source IDs and matched COSMOS2025 IDs, sky coordinates, multi-band photometry, photometric redshifts. The reduced imaging data will include the science mosaics used to construct the photometric catalog, together with the corresponding uncertainty maps. 

For the spectroscopic data, we will release the extracted NIRCam/F444W grism spectra for the sources analyzed in this work. 
The released spectra will include the coadded two-dimensional spectra and continuum-subtracted spectra, and the corresponding one-dimensional extractions with wavelength, flux, and uncertainty. 
These products will allow users to remeasure line fluxes, examine possible contamination, and test alternative line-identification methods.

The final line-emitter catalog will provide the spectroscopic redshifts, quality ranks, identified emission lines, observed line centers, line fluxes and uncertainties, broad-line AGN flags, possible contamination IDs, photometric redshifts, F444W magnitudes, and SED-derived physical properties. 
The quality ranks and contamination information should be used when defining statistical samples. The detailed description of this line-emitter catalog is presented in Section~\ref{subsec:line_cat}.

The catalog tables are provided in machine-readable format, while the imaging and spectroscopic data products will be released as FITS files. 
We also provide an interactive FITSMap interface
%\footnote{\textcolor{red}{xxxx, we need a domain name?}} 
for interactive visualization of the imaging and spectral data products. This web-based tool allows users to seamlessly navigate through the large-scale mosaics.
For each source in the photometric catalog, the interface integrates the multi-band SED alongside its corresponding photometric redshift PDF. 
For the identified line-emitters, we provide comprehensive diagnostic panels, including the 1D/2D spectra, the diagnostic summary generated by the LLM, etc.

\section{Discussion} \label{sec:discuss}
In this section, we discuss the reliability and limitations of the emission-line identification and redshift measurements in our catalog. We first compare the grism spectroscopic redshifts with photometric redshifts, LLM-assisted identifications, and previously published spectroscopic measurements to evaluate the overall accuracy and consistency of our methodology. We then investigate the main sources of uncertainty in the line-identification process, including ambiguous photometric redshifts, overlapping spectral traces in slitless spectroscopy, contamination from nearby bright sources, and complex multi-component systems. These analyses provide context for assessing the robustness of future statistical studies based on this sample.

\subsection{Redshift Comparison} \label{sec:z_compare}
\subsubsection{Photometric redshift vs. Grism spectroscopic redshift}

\begin{figure*}[!t]
\centering
\includegraphics[width=\textwidth]{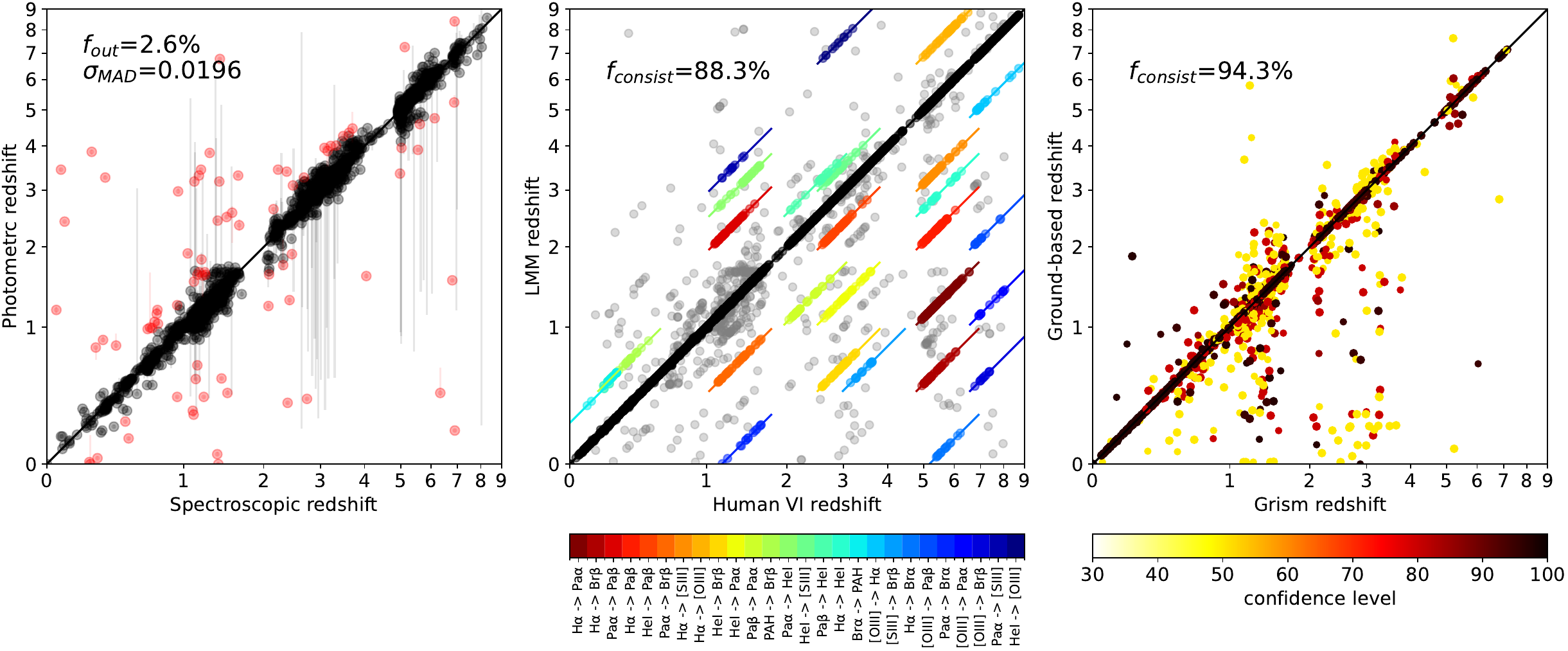}
\caption{
Comparison of redshift measurements for the confirmed line-emitter sample. 
\textit{Left:} Grism spectroscopic redshifts versus photometric redshifts. Outliers with $|z_{\mathrm{phot}} - z_{\mathrm{spec}}|/(1+z_{\mathrm{spec}}) > 0.15$ and $z_{\mathrm{spec}} > z_{\mathrm{phot,84}}$ or $z_{\mathrm{spec}} < z_{\mathrm{phot,16}}$ are highlighted in red. 
\textit{Middle:} Human-confirmed redshifts versus LLM results. Color-coded points labeled $A \rightarrow B$ indicate cases where a bright emission line initially identified as $A$ is reassigned to line $B$ based on human inspection. Colors correspond to the number of inconsistent sources (red: most, blue: fewest). Black points indicate consistent cases, and gray points represent other inconsistencies with small numbers. 
\textit{Right:} Grism spectroscopic redshifts versus spectroscopic redshifts from \cite{Khostovan_2025}. Points are colored by confidence level in \cite{Khostovan_2025}, and point size indicates rank in our sample (larger points correspond to higher rank).  \label{fig:compare_z}}
\end{figure*}

We compare the photometric redshifts with the grism spectroscopic redshifts for our sample, as shown in the left panel of Figure~\ref{fig:compare_z}. Both the photometric redshifts from the COSMOS2025 catalog and those derived in this work using the latest imaging data are considered during source confirmation. We thus present the photometric redshifts that are the closest to the final grism redshifts. The majority of sources exhibit a tight correlation along the one-to-one relation across the full redshift range, indicating an overall good performance of the photometric redshift estimates.
We only compare sources that have both a robust photometric redshift and a reliable grism redshift, defined as follows: the number of filters used to estimate the photometric redshift is greater than three, the photometric redshift probability distribution has a single peak, and the grism redshift has a rank greater than 1.
We quantify the scatter using the normalized median absolute deviation, obtaining $\sigma_{\mathrm{MAD}} = 0.0196$.
The outlier fraction, defined as sources with $|z_{\mathrm{phot}} - z_{\mathrm{spec}}|/(1+z_{\mathrm{spec}}) > 0.15$ and $z_{\mathrm{spec}} > z_{\mathrm{phot,84}}$ or $z_{\mathrm{spec}} < z_{\mathrm{phot,16}}$, is $f_{\mathrm{out}} = 2.6\%$.
We inspect a subset of the outliers and find that most of them have images contaminated by nearby bright sources, suggesting that their photometric redshifts are poorly constrained.

\subsubsection{LLM vs. human-confirmed spectroscopic redshift}

We then compare the redshifts provided by the LLM with the final human-validated redshifts in the middle panel of Figure~\ref{fig:compare_z}. The consistency rate is $88.3\%$, indicating that the LLM is a powerful tool that can effectively combine multiple sources of information to make redshift decisions in a manner comparable to the human inspection.

The color-coded points and short line segments represent cases in which a bright emission line initially identified by the LLM is reassigned to a different line after human inspection based on additional physical considerations. The colors are ordered according to the number of sources with inconsistent redshifts, and only cases with more than 10 sources are color-coded. These corrections are often motivated by the presence of additional fainter lines ($<3\sigma$) that are not formally detected but can still be recognized through visual inspection, such as the weaker components in commonly observed line pairs including \oiii\ $\lambda\lambda4959,5007$, [\ion{S}{3}]\ $\lambda\lambda9069,9531$, \ion{He}{1}\ $\lambda10830$, and H$\gamma$. 
Other subtle pieces of information that are difficult for the LLM to fully incorporate may also contribute to these corrections. For example, in systems containing pairs or clusters of galaxies, the redshift of the brightest source can often be robustly identified, allowing human inspectors to infer that nearby low-SNR companions are likely at the same redshift. In contrast, the current LLM framework generally treats these sources independently and may therefore assign inconsistent redshifts. These cases point out directions for future improvements and training of the LLM framework.

Among the inconsistent cases, the misclassification of \paa\ as \ha\ is the most common. This mainly reflects the large number of \paa\ emitters in our sample (see Figure~\ref{fig:all_spec}), together with the well-known degeneracy between the Lyman break and the $4000$~\AA\ break in photometric redshift fitting. 
From human inspection, these two identifications are mainly distinguished using morphology, imaging, and EW information: \paa\ emitters are typically extended low-redshift galaxies with optical detections and smaller EWs, while high-redshift \ha\ emitters are generally more compact and show larger EWs and dropout signatures in the bluer bands.

\subsubsection{Literature Ground-based vs. Grism-based Spectroscopic Redshifts}

Finally, we compare our spectroscopic redshifts with previously published spectroscopic measurements, as shown in the right panel of Figure~\ref{fig:compare_z}. We cross-match our catalog with the spectroscopic redshift compilation of \citet{Khostovan_2025}, resulting in 3,864 matched sources. Among these, we restrict the comparison to the 3,798 sources with \texttt{rank $>1$} in our catalog. 
We define consistent measurements as sources with $|\Delta z| < 0.015$, and find an overall consistency rate of $94.3\%$. We adopt this relatively loose threshold because some of our redshifts are determined from broad $3.3\ \mu$m PAH features, which can naturally introduce uncertainties of up to $|\Delta z| \sim 0.015$. Most inconsistent sources have relatively low confidence levels in the \citet{Khostovan_2025} catalog. 

We further inspect the discrepant cases and find that a subset exhibits complex morphologies, such as closely separated or partially overlapping systems. In these cases, the UV/optical and infrared emission lines may originate from different galaxies at different redshifts rather than from the same physical system. This effect is likely enhanced in some ground-based spectroscopic observations, where the relatively large fiber apertures can blend multiple nearby components into a single spectrum. This echoes the recently reported high rate of foreground contaminants for high-redshift galaxies resolved by JWST \citep{Liu_2026}.
We identify two interesting systems in which the morphologies in the bluer-band images differ significantly from those in the redder bands. At the redshifts reported by \citet{Khostovan_2025}, which are primarily based on DESI spectra, no plausible emission line is expected at the observed wavelength of the line detected in our grism data. We thus interpret these systems as a line-of-sight superposition of two galaxies at different redshifts. In these cases, the UV emission lines detected by DESI likely originate from the foreground galaxy, while the infrared emission detected in our WFSS data arises from a more obscured background galaxy. The detected infrared line is most likely \paa\ from the background source, leading to a different redshift assignment in our catalog.

\subsection{Ambiguities in Line Identification} \label{sec:ident_prob}

The accuracy of photometric redshifts is particularly important for the large fraction of sources that have only a single emission line detected.
As mentioned above, some sources exhibit multiple peaks in their photometric redshift probability distribution functions, or have images contaminated by nearby bright sources, which makes the photometric redshift estimates ambiguous.

In cases where multiple sources are aligned along the dispersion direction, this ambiguity becomes more severe.
Figure~\ref{fig:VI_tool} presents an example of a complex configuration, in which we successfully distinguish the \ion{He}{1} $\lambda10830$ and [\ion{S}{3}] $\lambda\lambda9069,9531$ emission lines from one source (ID~79356) from the \paa\ emission of another source (ID~79239), based on the presence of multiple lines.
The \paa\ line from ID~79239 appears as two features in the spectrum of ID~79356 due to its displaced locations in Modules~A and~B.
In other cases, particularly those involving a single emission line detected in only one module, such distinctions are difficult to make with the current data.
For these ambiguous cases, we list the possible contamination sources as \texttt{contm\_ID} in the catalog for both objects and assign them a maximum quality rank of 2.

%bright stars;
Another factor affecting the accuracy of our sample is contamination from bright stars or galaxies. Such contamination can not only bias the photometric redshift estimates of nearby sources, but also lead to the misidentification of other sources. We find that some detected emission lines originate from sources that are masked in the COSMOS-Web detection images.
In our source detection procedure, we adopt smaller and more precise PSF masks that vary with position across the field in Section~\ref{sec:image}, enabling the detection of additional sources near bright stars.  
In order to balance line-emitter completeness and the proportion of spurious detections, we retain the additional sources brighter than 24 mag. 
In the final catalog, these sources are assigned identification numbers starting from 1,000,000. As discussed in Section~\ref{sec:ident}, excluding these sources from the identification process could result in misassignment of emission lines in the spectra of other sources. Therefore, we retain them in the catalog, although the ranks are generally low because the nature of their emission lines remains mostly uncertain due to highly contaminated images and SEDs.
%Nevertheless, this catalog remains subject to a slight bias in its spatial distribution due to the loss of sources located directly behind bright stars and their diffraction spikes. 
For analyses that depend on accurate spatial information, such as studies of galaxy environments, local density measurements, or large-scale structure, we recommend that users apply the COSMOS-Web masks and exclude sources with ID $>1,000,000$.

\subsection{Redshift Confirmation with Additional Bands}  \label{sec:add_progrms}

Broader spectral coverage often enables the detection of additional emission lines, thereby improving the accuracy and robustness of redshift determinations. SAPPHIRES (GO\#6434; \citealt{Sun_2025_SAPPHIRES}; PI: Egami, E.) and POPPIES (GO\#5398; \citealt{Kartaltepe_2025_POPPIES}; PI: Kartaltepe, J.) are two primary parallel NIRCam WFSS programs that overlap with the COSMOS-3D field (see Figure~\ref{fig:cover}). 
SAPPHIRES provides F356W WFSS coverage over $\sim109$ arcmin$^2$ (10\%) of the COSMOS-3D field, while POPPIES provides F277W WFSS coverage over $\sim5.6$ arcmin$^2$ (0.5\%) and F322W2 WFSS coverage over $\sim28$ arcmin$^2$ (3 \%). For line emitters located within these regions, the additional WFSS filters provide independent spectral constraints that can break common single-line degeneracies and significantly improve the reliability of line identification.

In total, 726 of our F444W-selected line emitters also show emission lines in at least one additional WFSS filter. Using all available slitless spectral information, we re-determine the redshifts of these multiple-line emitters with the same pipeline described in Section~\ref{sec:sample}. Among them, 41 sources are found to have revised redshifts that differ from the initial F444W-based estimates, and four sources are identified as spurious line emitters. These results imply an estimated redshift accuracy rate of approximately 94\% for the full sample.
Inspection of the revised cases shows that most corrections arise from ambiguities in single-line identification. A common example is the detection of \ha\ in the additional WFSS spectra, which confirms that the line detected in the F444W spectra corresponds to \oiii\ rather than \ha. Such additional spectral constraints substantially reduce degeneracies between different line-identification solutions.

These results highlight one of the major strengths of wide-field JWST slitless spectroscopy. Although each individual WFSS filter covers only a limited wavelength range, combining multiple filters enables a much more secure reconstruction of galaxy redshifts and emission-line properties over a broad redshift interval. Future large-area multi-filter WFSS surveys will therefore provide substantially improved constraints on faint line-emitter populations and reduce the systematic uncertainties associated with single-line identification \citep[e.g.,][]{2024arXiv240812713S,Sun_2025_SAPPHIRES}.

\section{Summary} \label{sec:summary}

We present a systematic census of emission-line galaxies in the COSMOS field using JWST/NIRCam WFSS observations mainly from the COSMOS-3D program. Combining slitless spectroscopy with deep multi-band imaging, we construct a homogeneous spectroscopic catalog spanning $z \approx 0$--9. Our analysis framework enables robust emission-line detection, identification, and flux measurements in crowded slitless data.

We develop a multi-step identification procedure that combines template cross-correlation, photometric redshift constraints, and visual inspection. To further improve reliability and efficiency, we introduce a Siamese CNN to quantify morphological consistency between galaxy images and 2D spectra, and an LLM-assisted framework to emulate human inspection. This approach effectively mitigates contamination and source confusion, which are the main challenges in slitless spectroscopy.

The resulting catalog provides spectroscopic redshifts, emission-line measurements, and derived physical properties for a large sample of galaxies. The wide wavelength coverage of the F444W grism enables simultaneous detection of multiple emission lines, allowing a continuous census of galaxies across cosmic time. 
The redshift distribution shows distinct peaks associated with the different primary emission lines accessible over the observed wavelength range.
%The redshift distribution shows clear structures associated with different primary emission lines.

We construct composite spectra in several redshift bins and detect both strong and faint features, including hydrogen recombination lines, CO absorption bands, and Fe~\textsc{ii} lines. These results demonstrate the statistical power of the sample and provide constraints on the average physical conditions of galaxies.

This catalog provides a key reference dataset for studies of galaxy evolution, star formation, AGN activity, and large scale structure. It also highlights the potential of JWST/NIRCam WFSS for building large, unbiased spectroscopic samples over wide areas.
In future work, we will extend this framework to additional JWST WFSS surveys to build a larger and more uniform census of emission-line galaxies across different fields.

\section*{Acknowledgment}
We acknowledge support from the National Science Foundation of China (12225301).
We thank Fengwu Sun and Xiaojing Lin for sharing the WFSS data-reduction and automatic redshift-determination codes.
We thank Zechang Sun for helpful discussions that inspire the LLM framework. 
We acknowledge the use of the Doubao Large Language Model developed by ByteDance. 

\vspace{5mm}
\facilities{JWST (NIRCam), HST (ACS)}

\software{\texttt{astropy} \citep{astropy_2013},  
          \texttt{cigale} \citep{Boquien_2019},
          \texttt{emcee} \citep{Foreman-Mackey_2013},
          \texttt{JWST} \citep{jwst_bushouse_2023},
          \texttt{matplotlib} \citep{Hunter2007}, 
          \texttt{numpy} \citep{Harris2020}, 
          \texttt{photutils} \citep{photutils},
          \texttt{PSFEx} \citep{2011ASPC..442..435B},
          \texttt{pyneb} \citep{Pyneb_2015}
          \texttt{PyPHER} \citep{2016ascl.soft09022B},
          \texttt{SExtractor} \citep{Bertin_1996},
          }

\bibliography{sample631}{}
\bibliographystyle{aasjournalv7}

\end{document}